%% file: main.tex
\documentclass[sigplan,nonacm]{acmart}
\usepackage{tikz}
\usetikzlibrary{positioning,arrows.meta,fit,backgrounds,calc}
\usepackage{enumitem}
\usepackage{subcaption}
\usepackage{multirow}

\usepackage{listings}
\begin{document}

\title[ForgeTrain]{ForgeTrain: Forging Production-Grade Training Frameworks
via Harness-Driven AI Development}

\author{Qingfeng He}
\authornote{Equal contribution.}
\affiliation{%
  \institution{Tsinghua University}
  \city{Beijing}
  \country{China}
}

\author{Zhui Zhu}
\authornotemark[1]
\affiliation{%
  \institution{Tsinghua University}
  \city{Beijing}
  \country{China}
}

\author{Shangzhan Li}
\authornotemark[1]
\affiliation{%
  \institution{Harbin Institute of Technology}
  \city{Harbin}
  \country{China}
}

\author{Yaojian Chen}
\affiliation{%
  \institution{Tsinghua University}
  \city{Beijing}
  \country{China}
}

\author{Haojun Sun}
\affiliation{%
  \institution{ModelBest Inc.}
  \city{Beijing}
  \country{China}
}

\author{Xu Chen}
\affiliation{%
  \institution{ModelBest Inc.}
  \city{Beijing}
  \country{China}
}

\author{Leshan Li}
\affiliation{%
  \institution{Tsinghua University}
  \city{Beijing}
  \country{China}
}

\author{Yifei Shen}
\affiliation{%
  \institution{ModelBest Inc.}
  \city{Beijing}
  \country{China}
}

\author{Changjingxing Zhao}
\affiliation{%
  \institution{Tsinghua University}
  \city{Beijing}
  \country{China}
}

\author{Mengyuan Fan}
\affiliation{%
  \institution{Peking University}
  \city{Beijing}
  \country{China}
}

\author{Wenyu Guan}
\affiliation{%
  \institution{ModelBest Inc.}
  \city{Beijing}
  \country{China}
}

\author{Yiyun Zheng}
\affiliation{%
  \institution{ModelBest Inc.}
  \city{Beijing}
  \country{China}
}

\author{Yuxuan Zuo}
\affiliation{%
  \institution{ModelBest Inc.}
  \city{Beijing}
  \country{China}
}

\author{Zhen Li}
\affiliation{%
  \institution{ModelBest Inc.}
  \city{Beijing}
  \country{China}
}

\author{Zhenghang Luo}
\affiliation{%
  \institution{ModelBest Inc.}
  \city{Beijing}
  \country{China}
}

\author{Yuxuan Li}
\authornote{Corresponding authors.}
\affiliation{%
  \institution{Tsinghua University}
  \city{Beijing}
  \country{China}
}
\email{yxuanl1995@gmail.com}

\author{Xu Han}
\authornotemark[2]
\affiliation{%
  \institution{Tsinghua University}
  \city{Beijing}
  \country{China}
}
\email{han-xu@tsinghua.edu.cn}

\author{Zhiyuan Liu}
\authornotemark[2]
\affiliation{%
  \institution{Tsinghua University}
  \city{Beijing}
  \country{China}
}
\email{liuzy@tsinghua.edu.cn}

\input{sections/abstract}

\maketitle

\input{sections/introduction}
\input{sections/background}
\input{sections/method}
\input{sections/experiments}
\input{sections/related}
\input{sections/conclusion}

\bibliographystyle{ACM-Reference-Format}
\bibliography{references}

\appendix
\input{sections/appendix}

\end{document}

%% file: sections/abstract.tex
\begin{abstract}

%

Training large models still relies on general-purpose frameworks such as
Megatron-LM, whose generality tax constrains scenario-specific optimization
and adds runtime overhead through accumulated abstraction. AI code
generation reduces the cost of building a framework, and makes it affordable
to forge one per scenario.
We propose Forge Engineering: building a dedicated implementation
from scratch for each scenario and iteratively optimizing it toward peak
performance under correctness and usability constraints.
Dedicated implementations inherit no abstraction boundaries, so they
can integrate optimizations across the stack and reach a higher performance
ceiling.
We instantiate this
paradigm for training frameworks as \textbf{ForgeTrain}, which holds a trusted framework as a
golden reference and relaxes equivalence monotonically from Bit-for-Bit
to Surpass. Experiments across multiple model--hardware configurations show that ForgeTrain
consistently produces correct training engines and improves MFU over established
training frameworks by $4.7$--$33.2\%$. To our knowledge this is the first production-grade training
framework forged end-to-end by AI to match or surpass its human reference.

\end{abstract}

%% file: sections/introduction.tex
	\section{Introduction}\label{sec:intro}

Training a frontier model has become one of the most capital-intensive
activities in computing: a single frontier run already costs tens of
millions of dollars and is headed toward \$1B~\citep{cottier2024rising}, on
top of industry-wide AI-infrastructure capital on the order of \$700B in
2026~\citep{aicapex2026}. The training framework fixes the throughput, the
memory footprint, and the model-FLOPs utilization (MFU) of every run. The
arithmetic is blunt: against a \$700B capital base, raising training
performance by a mere 10\% is worth at least \$70B a year, and every point
of MFU a framework loses costs millions per run.

Capturing this value means driving each training scenario to its peak
performance. In practice, however, training a large model hardly avoids a
general-purpose engine such as
Megatron-LM~\citep{shoeybi2019megatron}. Its \texttt{megatron/} tree has
accumulated abstraction layers, configuration switches, and special-case
branches for every model, scale, and device it has ever served
(Section~\ref{sec:generality-tax}). Yet any one concrete model-and-hardware
pair exercises only a small slice of them. This generality tax has two components.
\emph{(1)~Suboptimality}: a single framework must serve all scenarios at
once, so the tensor sharding, memory layout, operator fusion, and
communication schedule that would be optimal for the case at hand are
constrained to stay viable across all the others; scenario-specific peak
performance is out of reach. \emph{(2)~Runtime
overhead}: the abstraction, indirection, and configuration dispatch
accumulated to host every scenario erode MFU at every step.
The heavier the inherited baggage, the greater the
suboptimality and runtime overhead.

The general framework was the rational response to an economic premise.
Writing code was expensive and slow, so software engineering maximized
reuse of what was already written, and extending a shared framework always
beat rewriting one per scenario. Coding agents, however, have driven the
cost of writing code sharply
down~\citep{vibetensor2026,anthropic_ccc2026}, and that changes the
calculus. The better move now is to forge a dedicated framework for each
scenario, fitted natively to its model and hardware, exempt from both
taxes at once.

\textbf{Prior AI attempts.} Agents have already been set to write entire
systems from scratch. VibeTensor~\citep{vibetensor2026}, the first fully
AI-generated deep-learning runtime, was synthesized without a reference
and runs $1.7$--$6.2\times$ slower than
PyTorch~\citep{paszke2019pytorch}, because no source of truth forces its
global behavior onto a production baseline. Claude's C
compiler~\citep{anthropic_ccc2026} was developed under differential
testing against GCC and grew to roughly $100$K lines that compile the
Linux kernel, yet it still delegates assembly and linking to GCC, and its
authors state that it is not production-ready.

These failures share a diagnosis. Building a complex system with agents
poses three essential problems. Performance must be pushed to the human
bar, correctness must hold throughout, and the development itself must
stay efficient. Each attempt secures at most two of the three.
VibeTensor anchors neither correctness nor performance, and the compiler
holds correctness yet stops short of production performance. A
production-grade system requires all three at once, and none has yet
emerged.

\textbf{ForgeTrain.} For training frameworks, we answer with
ForgeTrain\footnote{Code and forged engines:
\url{https://anonymous.4open.science/r/forgetrain_anon-FE80}.}
(Figure~\ref{fig:overview}), an autonomous agent loop that forges a
dedicated framework from an empty repository. The loop answers the three
problems in turn.
\begin{itemize}[leftmargin=*,topsep=2pt,itemsep=2pt]
  \item \textbf{How to reach peak performance?} The forged framework must
    approach and ultimately surpass the hand-tuned framework it replaces;
    anything less
    defeats the purpose of forging. The obstacle, however, is planning
    rather than capability: agents already forge kernels that
    rival vendor libraries; what a bare loop cannot do is lay
    out the path from empty repository to peak as a sequence of steps it can
    walk. ForgeTrain marks that path with Milestones and adjudicates
    each with a Gate: a real training run comparing training quality and
    throughput at once. Passing a Gate locks the gain into the baseline,
    ratcheting measured MFU toward the target round by round.
  \item \textbf{How to maintain correctness?} A faster framework that
    trains wrong is worthless, and correctness here is the hardest thing to
    adjudicate: it is a global property, surfacing only across whole
    training trajectories. ForgeTrain therefore disciplines the forge
    with a golden reference, a trusted framework such as
    PyTorch~\citep{paszke2019pytorch},
    Megatron-LM~\citep{shoeybi2019megatron}, or
    MindSpeed~\citep{mindspeed} held as ground truth, so
    correctness never needs to be
    specified from scratch. The crux is the order of
    correctness requirements: it first reproduces the reference's
    anchor, the artifacts of its training process, bit for bit,
    and relaxes the requirement only afterward. Bit-for-bit reproduction delivers a
    provably correct limit, not just a starting point. Forging under a bar that
    relaxes gradually from that limit turns debugging from conjecture over
    global behavior into pointwise comparison, which suppresses the risk of
    forging wrong.
  \item \textbf{How to forge efficiently?} The case for
    forging rests on the falling cost of writing code, and a loop that
    spends its rounds on anything but the engine forfeits that advantage.
    The obstacle is that the optimizations reaching toward the peak are
    multi-step and pass through worse intermediate states, so a bare loop,
    rewarded by the Gate one step at a time, stops at the nearest local
    optimum, while every round lost to a broken build or a drifted dataset
    is a round not spent on the engine at all. ForgeTrain therefore hardens
    the forging environment, freezing dependencies, build, data, and
    evaluation, and supplies a knowledge prior, a corpus of
    optimization experience recording which directions historically pay off
    and which intermediate costs are worth tolerating. The environment
    keeps every round on the engine; the prior lets the loop commit to a
    multi-step optimization before the Gate can reward it.
\end{itemize}
These three answers are ForgeTrain's method in sum: the performance and
correctness answers shape the protocol, the efficiency answer shapes the
environment, and together they hold the loop to one trajectory: reproduce
the reference exactly, then leave it behind.
The underlying paradigm, Forge Engineering, is to build a
dedicated implementation from scratch for each scenario and iteratively
optimize it toward peak performance under correctness and usability
constraints. ForgeTrain instantiates this paradigm for training frameworks.
The Harness is itself reused across scenarios: each new model-and-hardware
pair starts a fresh forge, drawing on the same construction knowledge and
executable evaluations. As coding costs fall,
this approach makes scenario-specific construction increasingly viable,
just as falling design cost once pushed hardware from general-purpose CPUs
toward domain-specific architectures~\citep{hennessy2019golden}.

\begin{figure}[t]
\centering
\includegraphics[width=0.83\linewidth]{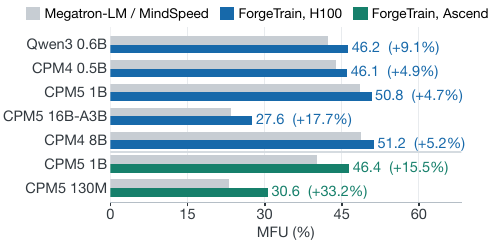}
\caption{Framework-level MFU of each ForgeEngine against its golden
reference; relative gain annotated.}
\label{fig:mfu-gain}
\end{figure}

\textbf{Results.} From an empty repository, ForgeTrain forges a dedicated
ForgeEngine for seven model--hardware settings, the same Harness
retargeted across two vendors: Megatron-LM on NVIDIA H100 and MindSpeed on
Ascend 910 NPUs. Every engine surpasses its reference, by
$4.7$--$33.2\%$ MFU at matched training quality
(Figure~\ref{fig:mfu-gain}), and its forged FlashAttention and GEMM kernels
run level with FlashAttention-3 and cuBLAS. Correctness is validated on the
three engines carried into long-horizon training, including MiniCPM4 0.5B,
MiniCPM5 1B, and MiniCPM5 130M, each holding loss parity with its reference through
the run and downstream parity with the reference-trained baseline.

Concretely, we make three contributions:
\begin{itemize}[leftmargin=*,topsep=2pt,itemsep=1pt]
  \item We propose \textbf{Forge Engineering}, a paradigm for
    building systems software from scratch for each scenario and iteratively
    optimizing the implementation toward peak performance under correctness
    and usability constraints. A reusable Harness supports this process
    through construction knowledge and executable evaluations.
  \item We propose \textbf{ForgeTrain}, which instantiates
    Forge Engineering for training frameworks. A trusted framework serves
    as the golden reference: the loop first reproduces its training
    artifacts bit for bit, then relaxes the requirement stage by stage to
    training-quality parity, so the forged engine first equals the
    reference and then beats it. Milestones and the Gate drive the
    throughput gains; long-run validation checks correctness after each
    relaxation; a hardened environment and a knowledge prior keep the loop
    efficient.
  \item We deliver \textbf{ForgeEngine}, the per-scenario engines ForgeTrain
    forges: they match their golden reference on training quality and
    surpass it by $4.7$--$33.2\%$ in throughput, and we run MiniCPM4 0.5B's
    decay phase with its engine to downstream-evaluation parity with a
    Megatron-trained baseline. To our knowledge this is the first
    production-grade training framework forged end-to-end by AI to match or
    surpass its human reference.
\end{itemize}

%% file: sections/background.tex
\section{Background}\label{sec:background}

\subsection{Generality Tax and Forge Engineering}
\label{sec:generality-tax}

Megatron-LM~\citep{shoeybi2019megatron} makes the generality tax concrete. The
framework's source code has grown over time to accommodate new models, scales,
and devices, and has accumulated layers of abstraction, configuration switches,
and special-case branches. Any one concrete model-and-hardware pair, however,
exercises only a small slice of them. Two measurements make the tax visible.
First,
suboptimality: across the $128{,}275$ lines of its \texttt{megatron} directory
in \texttt{core\_v0.15.0}, roughly one line in every $134$ is a configuration
branch on \texttt{args}, \texttt{config}, or \texttt{self.config}. The tensor
sharding, memory layout, operator fusion, and communication schedule that would
be optimal for the case at hand are constrained to stay viable for all the
scenarios the framework serves. Second, runtime overhead: the
abstraction built to host those scenarios is equally deep. Whereas a dedicated
engine can invoke the kernel directly, a single QKV projection traverses seven
Python module boundaries before reaching the GEMM, including \texttt{GPTModel},
the block, layer, and attention modules, a Transformer Engine wrapper layer,
an \texttt{autograd.Function}, and the \texttt{aten} dispatcher
(Appendix~\ref{app:ascend-opt-catalog}).

Communication--computation overlap illustrates the suboptimality this
generality imposes. Megatron-LM's tensor-parallel overlap relies on the
process-wide environment variable
\texttt{CUDA\_\allowbreak DEVICE\_\allowbreak MAX\_\allowbreak CONNECTIONS} being set to~$1$: with a single
hardware queue, the all-gather issued ahead of the GEMM is guaranteed to be
scheduled first, and the framework asserts this value whenever tensor or
context parallelism is enabled. Expert-parallel overlap for MoE models needs
a value larger than~$1$, so that dispatch communication and
expert computation can actually run concurrently. When a MoE model uses both
forms of parallelism, the framework can only warn that the user should
``set \texttt{CUDA\_DEVICE\_MAX\_CONNECTIONS} to 1 or 32, depending on which
parallelization you want to prioritize.'' Both overlaps are implemented and
each is optimal in isolation, yet a single global knob that must remain valid
for every scenario forces the combined case to forfeit one of them. A dedicated
implementation for a fixed workload can instead express the required ordering
through explicit stream and event dependencies and obtain both overlaps at
once.

\begin{figure}[t]
\centering
\includegraphics[width=\linewidth]{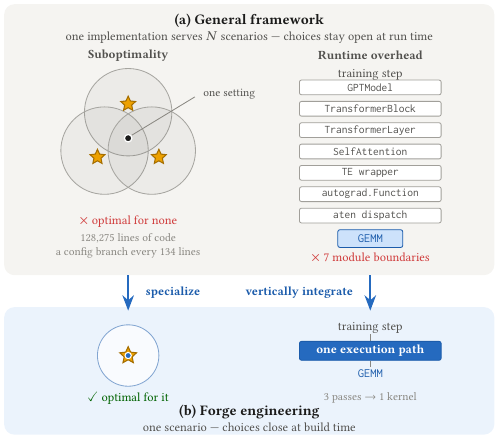}
\caption{The two forms of the generality tax (top) and the two freedoms that
answer them one for one (bottom). $\star$~marks each scenario's own optimal
setting.}
\label{fig:generality-tax}
\end{figure}

Forge Engineering
uses this freedom to build a dedicated implementation from scratch and
iteratively optimize it toward peak performance under correctness and
usability constraints. The deployment scenario specifies the workload,
hardware, and operational requirements. Iteration proceeds through
implementation, execution, evaluation, and revision; each accepted candidate
must satisfy the applicable correctness and usability checks. A trusted
reference anchors the behavior to preserve. The Harness supports this process
with reusable construction knowledge and executable evaluations, allowing
the same process to guide separate implementations across scenarios.
ForgeTrain instantiates this paradigm for training infrastructure.

\subsection{Coding Agents and Harness}\label{sec:agents-harness}

A coding agent is a language model that acts through tool calls: it inspects
a repository, edits files, and invokes build and test commands. These calls
become a development process only under an agent harness, or scaffold,
which supplies the execution loop, tools, context management, and instructions
that enable a model to act. Restrained by that loop, an agent revises its
implementation using execution feedback, and can take on complex, often
multi-file software-engineering tasks that take human professionals hours or
days, as SWE-Bench Pro evaluates~\citep{swebenchpro2025}; it can even forge
low-level operators, with Sakana's CUDA Engineer synthesizing thousands of
test-verified CUDA kernels~\citep{sakana_cuda2025}.

The general harness turns the model into a sustained developer rather than a
single-turn responder: across context windows the agent must recover project
state, identify unfinished work, preserve progress, and verify changes.
Anthropic's harness for long-running coding agents supports exactly this,
where an initializer prepares the environment and feature list while later
sessions make incremental changes, test them, and leave progress records and
Git commits for subsequent sessions~\citep{anthropic_longrunning2025}.
Harnesses are, however, not one-size-fits-all:
task-specific harnesses adapt the loop and its evaluation to a domain.

For training infrastructure, correctness is a global property of an entire
optimization run rather than of any local unit test, and it must be judged
alongside throughput. ForgeTrain therefore provides a task-specific Harness
that organizes construction through Milestones and executable Gates,
progressing from exact reproduction of reference anchors to performance
optimization under training-quality constraints. Section~\ref{sec:method}
describes the protocol.

%% file: sections/method.tex
\section{ForgeTrain: Method}\label{sec:method}


\subsection{Overview: The Agent Loop and the Golden Reference}\label{sec:method-overview}

\begin{figure*}[t]
\centering
\includegraphics[width=\linewidth]{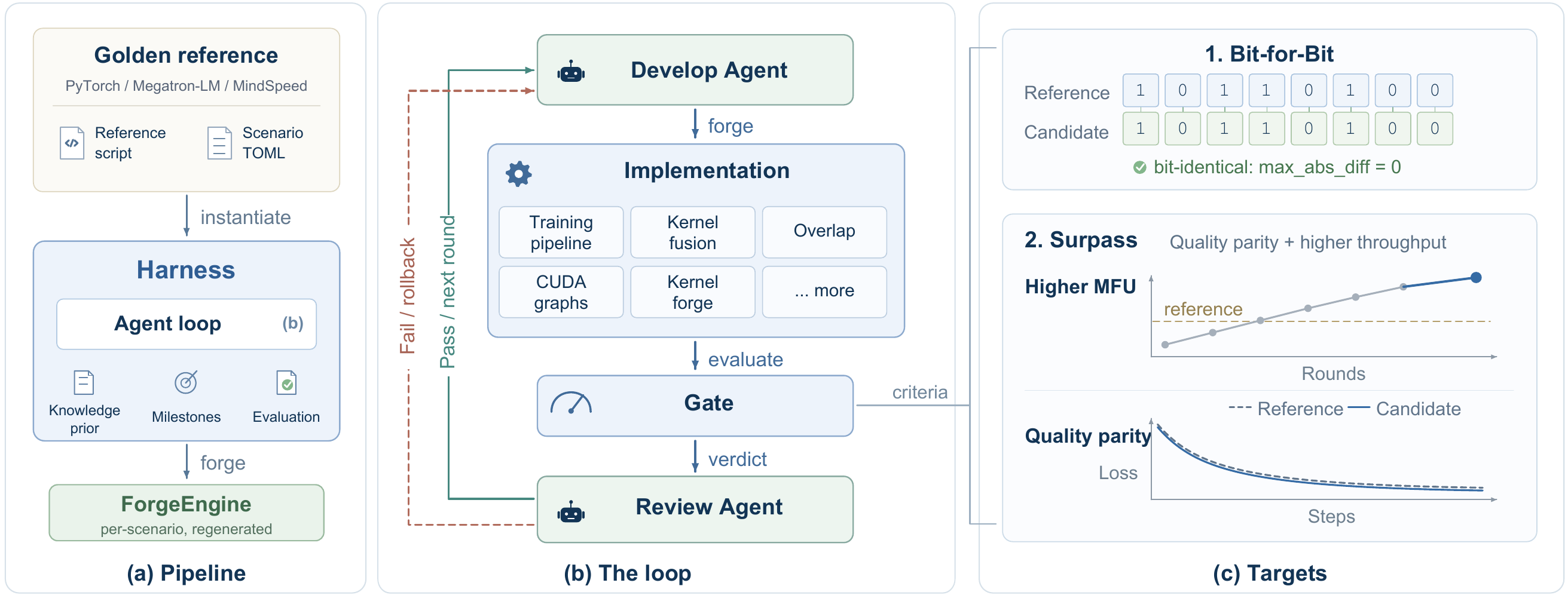}
\caption{\textbf{ForgeTrain at a glance.}
\textbf{(a)}~The Harness turns a golden reference and a thin per-scenario
configuration into a ForgeEngine; \textbf{(b)}~the forge--evaluate--verdict
loop that builds it; \textbf{(c)}~the targets the Gate enforces: bit-for-bit
identity, then higher MFU with a loss curve that tracks the baseline.}
\label{fig:overview}
\end{figure*}

ForgeTrain turns one training scenario into a framework built for that
scenario alone (Figure~\ref{fig:overview}a). Its inputs are a golden
reference (PyTorch~\citep{paszke2019pytorch},
Megatron-LM~\citep{shoeybi2019megatron}, or
MindSpeed~\citep{mindspeed}) and a thin per-scenario
configuration: a reference script that drives the golden reference and a
single TOML file describing the target model and hardware. The Harness, the
one component reused across every forge, guides and constrains a coding
agent through the build, and what the agent yields is a dedicated
ForgeEngine fitted to that scenario and to nothing else. No human stands
inside the loop: once started, the agent generates, runs, and revises the
framework on its own.

Prior to the forging loop, the agent instantiates a
scenario-specific Harness. Working from the reference script and the TOML, it
runs the golden reference once and captures the artifacts of that run as the
bit-for-bit anchors the engine will be held to, and it writes the concrete
Milestone and Gate scripts for the target model, hardware, and parallel
configuration. Once forging begins, these scripts are frozen on the Harness
side, out of the develop agent's reach (\S\ref{sec:gate}). Forging then
unfolds in two distinct stages. The first demands an exact match at every
anchor: the agent loop establishes a bit-for-bit implementation of the
training behavior the golden reference specifies. The second relaxes that
demand to training-quality parity, and the agent uses the resulting
reference-anchored evidence to guide performance optimizations whose
numerical paths may differ (Table~\ref{tab:stages}).

\begin{table*}[t]
\centering
\small
\caption{The three phases of a forge.}
\label{tab:stages}
\begin{tabular}{@{}llll@{}}
\toprule
\textbf{Phase} & \textbf{Equivalence demanded} & \textbf{Freedom granted} & \textbf{Success criterion} \\
\midrule
0: Instantiation & ---                      & ---                        & Milestones, Gate scripts, and bit-for-bit anchors in place \\
1: Bit-for-Bit   & Bit-level identity       & None beyond the reference  & Exact match at every anchor \\
2: Surpass       & Training-quality parity  & Restructure, fuse, rewrite & Quality parity, higher throughput \\
\bottomrule
\end{tabular}
\end{table*}

The objective is realized through repeated forging rounds
(Figure~\ref{fig:overview}b): the develop agent proposes a change, the
Harness evaluates it against the active milestone, and the review agent
verifies the implementation and the evaluation path. The measured outcome
then determines the context of the next round.

A training framework decomposes into two layers, and the loop proceeds
through both. The framework layer holds the pipeline, communication,
optimizer-state, and scheduling code that treats each operator as a unit;
beneath it, the operator layer holds the kernels the pipeline
invokes. Both layers are forged under the same discipline, and the agent uses
profiling evidence to select operators for further optimization after the
framework layer is established. The forged engines along these two layers
are analyzed in \S\ref{sec:throughput}--\S\ref{sec:operators}.

The following subsections describe the performance protocol
(\S\ref{sec:gate}), the correctness chain (\S\ref{sec:correctness}),
and the mechanisms for efficient forging (\S\ref{sec:wiki}).

\subsection{Peak Performance: Milestones and the Gate}\label{sec:gate}

For a coding agent, reaching peak performance is primarily a planning
problem rather than an implementation problem. Frontier agents can already
implement many local optimizations, including specialized kernels
\citep{kernelevolve2026,sakana_cuda2025} and large system components
\citep{anthropic_ccc2026}, but local capability does not determine which
dependent changes to attempt or how to organize them into a complete path from
an empty repository to peak throughput. ForgeTrain addresses this gap with
Milestones and Gates.

\textbf{Milestone decomposition.}\label{sec:milestone} A Milestone is an
intermediate target verified against the golden reference, and the
decomposition is recursive: a long-range target splits into sub-goals, each
sub-goal into shorter steps, until every step is short enough for the
develop agent to complete and the reference to check. The decomposition is designed for the capabilities and dependencies
of the target model, hardware, and parallel configuration, so different
scenarios may expose different intermediate milestones. It is written at
instantiation (\S\ref{sec:method-overview}): the MiniCPM4 8B scenario, for
instance, carries a Milestone in which the agent implements tensor
parallelism on its own, whereas the 0.5B scenario needs no tensor parallelism
and carries a data-parallelism Milestone in its place. During reference
reproduction, a Milestone can be an operator-level checkpoint; during
performance optimization, it can be an intermediate performance objective.
Milestones are therefore the finest granularity at which the loop advances,
and the decomposition fixes both how the work proceeds and when each piece of
it may be admitted. The Forge Loop is precisely this progression:
round by round, the agent advances through the sequence of Milestones.

\textbf{The Gate.} A Gate is the Harness-controlled verdict on
whether a candidate implementation satisfies a prescribed condition, measured
in a real training run. Milestones set the targets; Gates decide when a target
counts as reached, so every Milestone is closed by a Gate that checks its
advancement condition, such as training quality within bounds and throughput
at the mark for a performance Milestone. Over a multi-round performance
optimization the Gate serves three functions. First, measured positioning: every verdict returns the
candidate's measured quality and throughput, so the develop agent judges its
progress from data rather than from an unverified performance estimate.
Second, actionable failure: a rejection states whether the problem is
training quality out of bounds or speedup short of the mark, which gives the
next round a concrete direction. Third, safe accumulation: optimizations
that pass are written into the current baseline and become the starting
point of later rounds, while failed changes are rolled back whole, so
performance already won is preserved without carrying a failed exploration
into the state that follows. The performance target is set above the current
implementation and no single optimization reaches it, so MFU rises through
successive measurement and admission (Figure~\ref{fig:overview}c).

A capable agent may try to hack a Milestone's Gate rather than earn
it, so the Harness enforces a privilege split: the agent can only edit the
candidate implementation, while Gate conditions, run shapes, and reference-side
artifacts belong to the Harness and stay out of its reach. The review agent
additionally audits for proxy execution, fabricated metrics, altered workloads,
or weakened evaluation logic. Progress can therefore come only from candidate
changes that pass the prescribed Gates.

\subsection{Correctness: The Equivalence Chain and Long-Run Validation}\label{sec:correctness}

Every step of performance presupposes correctness, and at this scale a
training framework's correctness is a global property. Collective communication,
optimizer-state management, mixed-precision policy, and scheduling go wrong
in ways that surface only across an entire training trajectory, never in a
local unit test. The requirement is moreover double: the framework
must reproduce the golden reference exactly to be trustworthy, yet diverge
from it to be faster. No fixed target is at once ``equal to'' and ``better
than'' the reference. The equivalence chain resolves the tension by
separating the two requirements in time, descending a lattice of
equivalence relations from bit-level identity to training-quality parity;
relaxation is one-way and happens at stage granularity, which is one level
coarser than the Milestones of \S\ref{sec:gate}. The order carries the
argument: before any optimization begins, the engine agrees with the
reference to the bit at every anchor captured during instantiation, so any
error a later rewrite introduces localizes to a concrete deviation at a
concrete tensor rather than a guess about global behavior. This progression
has not, to our knowledge, been articulated as a methodology: prior
reference-anchored work uses the reference only for differential testing,
and reference-free synthesis has no reference to tighten against.

\textbf{Bit-for-Bit: zero-tolerance adjudication.} The anchors are the
artifacts of the golden reference's own training run on the target scenario,
captured at instantiation as machine-checkable assertions: the activations,
the gradients, the optimizer states, the loss-scaling trajectory, and the
collective-communication pattern, together with the invariants over their
ordering (a given all-reduce must complete before the corresponding
parameter update). They are taken at per-operator granularity, at exact
numerical values, and over the edge cases production must survive but unit
tests omit (gradient overflow and loss-scale backoff, checkpoint
save-and-restore across a parallelism change, the first step after
resumption); weaker anchors admit an implementation correct under normal
inputs yet silently wrong in production. The anchor set is discovered by the
agent's own instrumentation rather than curated by hand. The forged framework
must then clear those anchors with exact equivalence: given identical inputs
and seeds, every anchored tensor satisfies $\texttt{max\_abs\_diff}=0$. One
deviating bit at any anchor is failure. We demand this over the usual
loss-curve agreement because it rules out agreement for the wrong reasons:
canceling errors, an unstable kernel passing on a benign input, and a wrong
reduction order hidden inside an acceptable loss can each fool a curve-level
comparison, but none survives zero tolerance. Any deviation is reported as a
localized defect to be fixed, never an optimization to be kept. Our
single-node, eight-GPU ForgeEngine build clears
the full anchor set at roughly $80\%$ of Megatron-LM's throughput: slower,
but provably computing the reference's result.

\textbf{Long-run validation: holding parity after the relaxation.} The
Surpass stage then drops the equivalence one notch to training-quality
parity, and the develop agent may restructure freely, fusing operators,
reordering reductions, overlapping communication, and rewriting recomputation
policies. These moves break the bit-level anchors, so the enforcement of
correctness passes to the real training run behind every gate
(\S\ref{sec:gate}), read through sensitive indicators: the loss trajectory
must coincide with the baseline within run-to-run variation, downstream
evaluation must hold under an identical fine-tuning recipe, and
gradient-norm statistics, which are more sensitive to numerical perturbation
than the loss, expose a deviation before it reaches the loss, pinning the
error to the round that introduced it. The runs must be long because a
framework's errors, such as a wrong reduction order or a subtle
mixed-precision slip, are invisible in a single step and accumulate only over
a trajectory. And the
criterion is valid only in this order: first prove the engine computes the
quantities the reference intends, then allow it to compute them by a
different numerical route.

\subsection{Forging Efficiency: The Environment and the Knowledge Prior}\label{sec:wiki}

The protocol above answers what each step should do and whether it was done
right; forging efficiency decides how many rounds those steps take. A loop
that spends rounds on anything but the artifact erodes the near-zero
forging cost the enterprise rests on, and the waste comes from two places,
friction in the environment and detours in the search. ForgeTrain answers
each with one design.

\textbf{A hardened forging environment.} The first waste has nothing to do
with the artifact: a dependency that fails to install, a build that breaks
intermittently, an evaluation entry point that does not reproduce. Every
round the develop agent loses to such accidents is pure loss, and the
accidents cascade: one mis-installed dependency contaminates every verdict
after it. The Harness therefore hardens the environment up front:
dependencies, build, data, and evaluation entry points are frozen at
instantiation (\S\ref{sec:method-overview}), before the loop starts, and the
fully codified evaluation of \S\ref{sec:gate} pays a
second dividend here, since a verdict that cannot be argued with can also be
triggered cheaply at any moment. Every round the develop agent spends
therefore goes into the engine itself rather than into the surrounding
infrastructure.

\textbf{The knowledge prior.} The second waste comes from the search itself.
The rewrites that approach the peak are multi-step and their intermediate
states are worse: a fusion pays off only after the memory layout is also
changed; a reduction reorder helps only once communication is overlapped. An
unaided search stops at the nearest local optimum, because the gain lies
past a run of worse candidates. The Gate admits each of those stopped steps
as it should, but it does not supply the next one. The limit here is not
capability: a frontier coding agent can write each of these rewrites, but it
will not commit to a sequence whose early steps measure worse. The
knowledge prior addresses this by raising the search's effective
learning rate. It is a durable corpus of optimization experience the develop
agent consults, recording which aggressive directions pay off in a given
regime, which intermediate costs are worth tolerating, and which changes
compound. Consulting it leads the develop agent to accept a run of worse
candidates and reach the optimum beyond them. Every such attempt still
passes the Gate, so the prior widens the search without weakening
correctness: the prior determines what the loop attempts, the Gate what is
admitted. The prior itself never writes code, and it lives in the
Harness rather than in the forged framework, leaving the
zero-human-written-code
discipline intact.

The three groups of mechanisms are independent of one another: each governs
a different property of the forge, and none weakens the guarantees the other
two provide. Together they are what makes the match-then-surpass progression
against the golden reference executable by an imperfect agent.

%% file: sections/experiments.tex
\section{Experiments}\label{sec:experiments}

\subsection{Setup}\label{sec:setup}

We evaluate ForgeTrain through three experiments. The first measures
performance by comparing the MFU reached during forging with established
training frameworks. The second checks correctness by carrying the forged engines through
end-to-end forging and production training. The third evaluates the
usefulness of the Harness by conducting a forging task after removing selected
Harness components. Sections~\ref{sec:mfu},
\ref{sec:validation}, and \ref{sec:ablation} describe these experiments in
this order.

\textbf{Hardware and software.} We conduct the experiments on two platforms:
NVIDIA H100 GPUs and Huawei Ascend 910 NPUs. On H100, we use PyTorch~2.8.0
with CUDA~12.9, cuDNN~9.10.02~\citep{chetlur2014cudnn}, and NCCL~2.27.3.
The forged H100 engines use Transformer Engine~2.4~\citep{nvidia_te}, and
the hand-written operators use CUTLASS~4.5.0~\citep{nvidia_cutlass}. On
Ascend, we use PyTorch~2.7.1+cpu with CANN~8.5.0 and \texttt{torch\_npu}~2.7.1. All experiments use a
sequence length of $4096$ and the mixed-precision settings specified for the
corresponding platform. Model architectures are given in
Appendix~\ref{app:models} for the H100 settings and
Appendix~\ref{app:ascend-models} for the Ascend settings; the parallel degrees
and batch sizes for each setting are in Table~\ref{tab:mfu-checkpoints}.

\textbf{Performance experiment.} We forge training frameworks for seven
model--hardware settings. On H100, we use the official Claude Opus~5 agent
(1M-token context window) with ForgeTrain to forge Qwen3 0.6B, MiniCPM4 0.5B, MiniCPM5 1B, MiniCPM5 16B A3B,
and MiniCPM4 8B. We compare their MFU with Megatron-LM
0.15~\citep{shoeybi2019megatron}. On Ascend, we forge MiniCPM5 1B and MiniCPM5
130M and compare their MFU with their golden references, MindSpeed on Megatron
core~0.12.1 and PyTorch
respectively. MiniCPM4 0.5B and MiniCPM5 1B on H100 and MiniCPM5 1B on Ascend each carry
a one-layer MTP (Eagle) head; the other settings train without MTP.
Table~\ref{tab:mfu-checkpoints} collects the framework configurations and
the corresponding MFU trajectories.

\textbf{Correctness experiment.} On H100, we use Claude Opus~4.6 to forge a
MiniCPM4 0.5B engine, with Megatron-LM as the golden reference. We run the
forged engine through the decay phase of pretraining, then fine-tune the
resulting checkpoint with Megatron-LM and evaluate it on downstream
benchmarks. On Ascend, we use Opus~4.6 with a 1M-token context window to forge the
MiniCPM5 1B engine (with a one-layer MTP head) with MindSpeed as the golden
reference, and GLM-5.2 with a
1M-token context window to forge the MiniCPM5 130M engine with PyTorch as the
golden reference. We run decay training with the 1B engine and a complete
stable--decay--SFT training pipeline with the 130M engine, and evaluate the
resulting checkpoints of both on downstream benchmarks.

\textbf{Ablation experiment.} We use Claude Opus~5 with a 1M-token context
window for two ablations on H100. We remove the Harness entirely for MiniCPM5 16B A3B, and remove the quality
constraints while retaining the Milestones for MiniCPM4 0.5B. We use the
corresponding complete-Harness runs in Section~\ref{sec:mfu} as comparisons.

\textbf{Metric.} We report throughput as model-FLOPs utilization (MFU), computed
from an exact per-operator FLOP count rather than the usual closed-form
approximation (the full expression is given in Appendix~\ref{app:flops}). The
same expression and the same peak-times-devices denominator are used for the
forged engine and for the baseline.

\subsection{Main MFU Results}\label{sec:mfu}

We first report framework-level performance across seven model--hardware
settings, then present the kernels selected and forged by the agent during
the MiniCPM4 0.5B run. We conclude with an analysis of the framework and
kernel implementations.

\subsubsection{Framework-Level Forging}\label{sec:throughput}

ForgeTrain surpasses the corresponding Megatron-LM or MindSpeed baseline
across all seven model--hardware settings (Tables~\ref{tab:mfu-checkpoints}),
with relative MFU gains of $4.7\%$--$33.2\%$. The dense H100 models gain
approximately $5\%$--$9\%$, while MiniCPM5 16B A3B and the Ascend models show larger
improvements of about $16\%$--$33\%$. Table~\ref{tab:mfu-checkpoints} shows how
each setting reaches that result. Forging cost has two readings: the round at
which the engine first passes its baseline,
roughly $12$--$20$ on H100 and $3$--$5$ on Ascend, and the round at which it
peaks, the last checkpoint $k$ reported for each setting in
Table~\ref{tab:mfu-checkpoints}.
These results demonstrate that
scenario-specific forging can improve performance across both hardware
platforms.

\begin{table*}[t]
\centering
\small
\setlength{\tabcolsep}{3pt}
\caption{Framework configurations and MFU trajectories during forging. Par.
gives the data (DP), tensor (TP), and expert (EP) parallel degrees; GBS and
MBS are global and micro-batch size; per-model details are in
Appendix~\ref{app:models}. $k$ indexes a setting's accepted forging rounds
(Appendix~\ref{app:framework-opt}); a slash marks a run that ended before that
checkpoint. MiniCPM4 0.5B and MiniCPM5 1B on
H100 and MiniCPM5 1B on Ascend carry a one-layer MTP (Eagle) head; the remaining settings
train without MTP.}
\label{tab:mfu-checkpoints}
\begin{tabular}{@{}lccrcccccccccc cc@{}}
\toprule
Setting & Par. & GBS & MBS & Baseline (\%) & \multicolumn{9}{c}{MFU (\%) at checkpoint $k$} & Final (\%) & Gain \\
\cmidrule(lr){6-14}
\multicolumn{5}{l}{\textit{H100 (Megatron-LM)}} & $k{=}4$ & $k{=}8$ & $k{=}12$ & $k{=}16$ & $k{=}20$ & $k{=}24$ & $k{=}28$ & $k{=}32$ & $k{=}36$ & & \\
\midrule
Qwen3-0.6B & DP2 & 80 & 10 & $42.4$ & $36.2$ & $36.3$ & $44.3$ & $45.4$ & $46.3$ & $45.6$ & $46.8$ & /   & /   & $46.2$ & $+9.1\%$ \\
MiniCPM4-0.5B & DP2 & 80 & 10 & $43.9$ & $25.0$ & $36.9$ & $44.2$ & $45.2$ & $46.1$ & /   & /   & /   & /   & $46.1$ & $+4.9\%$ \\
MiniCPM5-1B & DP2 & 80 & 4 & $48.5$ & $43.6$ & $44.9$ & /   & /   & /   & /   & /   & /   & /   & $50.8$ & $+4.7\%$ \\
MiniCPM5-16B-A3B & EP8 & 32 & 4 & $23.5$ & $18.4$ & $22.2$ & $23.0$ & $23.7$ & $24.7$ & $24.1$ & $25.9$ & $26.7$ & $27.6$ & $27.6$ & $+17.7\%$ \\
MiniCPM4-8B & DP4\,TP2 & 32 & 2 & $48.7$ & $23.1$ & $44.5$ & $46.1$ & $47.2$ & $50.9$ & /   & /   & /   & /   & $51.2$ & $+5.2\%$ \\
\midrule
\multicolumn{5}{l}{\textit{Ascend 910 (MindSpeed)}} & $k{=}1$ & $k{=}3$ & $k{=}5$ & $k{=}7$ & $k{=}9$ & $k{=}11$ & $k{=}13$ & $k{=}15$ & $k{=}17$ & & \\
\midrule
MiniCPM5-1B & DP2 & 80 & 2 & $40.2$ & $30.1$ & $44.8$ & $46.3$ & $46.4$ & /   & /   & /   & /   & /   & $46.4$ & $+15.5\%$ \\
MiniCPM5-130M & DP2 & 80 & 5 & $23.0$ & $10.0$ & $22.6$ & $27.1$ & $29.4$ & $29.9$ & $30.4$ & $30.6$ & /   & /   & $30.6$ & $+33.2\%$ \\
\bottomrule
\end{tabular}
\end{table*}

\subsubsection{Operator-Level Forging}\label{sec:operators}

This section analyzes the operator forge carried out as part of the H100
MiniCPM4 0.5B correctness experiment. The framework and operator results
reported here come from the same no-MTP MiniCPM4 0.5B forge. The agent first
profiles the training step to identify the operators that account for the
largest runtime cost.

The profile (Table~\ref{tab:opbudget}) leads the agent to select five GEMM
variants and the attention core for further forging. The GEMMs cover the QKV
projection, output projection, and the two FFN projections. For attention, the
agent forges both the forward and backward kernels. The optimization families
these forgings instantiate are cataloged in Appendix~\ref{app:opt-catalog}.

For each selected operator, the agent writes a specialized implementation for
the shapes issued by the MiniCPM4 0.5B training step. Each implementation is
checked against the corresponding reference computation before it is included
in the forged engine. The resulting kernels are then compared with the
corresponding vendor implementations at the target shapes
(Table~\ref{tab:kernels}).

\begin{table}[t]
\centering
\small
\caption{Operator time budget of a steady-state training step (bfloat16;
forward${+}$backward${+}$optimizer, averaged over $5$ profiled steps). Shares
are of GPU compute time, communication excluded; the GEMM row covers forward,
dgrad, and wgrad.}
\label{tab:opbudget}
\begin{tabular}{@{}lc@{}}
\toprule
\textbf{Operator class} & \textbf{Share of step} \\
\midrule
GEMM (\texttt{qkv}/\texttt{o}/\texttt{fc1}/\texttt{fc2}/\texttt{lm-head}) & $55.9\%$ \\
Attention core (cuDNN FlashAttention fwd${+}$bwd) & $16.6\%$ \\
Elementwise (RMSNorm, RoPE, activation, Adam) & $24.4\%$ \\
Other (Triton concat, sort) & $3.1\%$ \\
\bottomrule
\end{tabular}
\end{table}

\begin{table}[t]
\centering
\small
\caption{Forged-kernel throughput (MFU, \%) at the 0.5B engine's own shapes
(H100, bfloat16; weight gradients accumulated in FP32). FA-3 and FA-4 are
FlashAttention-3~\citep{shah2024flashattention3} and
FlashAttention-4~\citep{zadouri2026flashattention4}; TE is Transformer Engine.
``n/a'' marks passes the fused \texttt{qkv}/output backward kernel does not
expose separately. Bold marks the best value in each comparison (ties within
$0.1$\,pp both bold).}
\label{tab:kernels}
\begin{tabular}{@{}lccccc@{}}
\multicolumn{6}{c}{\textbf{(a) Attention}}\\
\toprule
\textbf{Direction} & \textbf{Forged} & \textbf{cuDNN} & \textbf{FA-3} & \textbf{FA-4} & \textbf{TE} \\
\midrule
Forward  & 42.6 & \textbf{47.8} & 44.3 & 43.7 & 39.0 \\
Backward & \textbf{45.7} & 35.0 & 43.4 & 43.3 & 36.2 \\
\bottomrule
\end{tabular}
\vspace{1.2ex}

\begin{tabular}{@{}lcccc@{}}
\multicolumn{5}{c}{\textbf{(b) GEMM (forged / cuBLAS)}}\\
\toprule
\textbf{Operator} & \textbf{Forward} & \textbf{Backward} & \textbf{dgrad} & \textbf{wgrad} \\
\midrule
\texttt{qkv}    & 69.7/\textbf{69.9} & 68.4/\textbf{73.7} & \textbf{72.9}/\textbf{72.8} & 62.4/\textbf{71.6} \\
\texttt{attn-out} & \textbf{70.8}/69.8 & \textbf{73.0}/71.5 & \textbf{69.9}/\textbf{69.9} & \textbf{72.7}/69.3 \\
\texttt{fc1}    & \textbf{77.1}/74.2 & \textbf{81.4}/\textbf{81.5} & 81.9/\textbf{82.1} & \textbf{83.3}/81.1 \\
\texttt{fc2}    & 80.2/\textbf{80.5} & \textbf{79.9}/78.9 & \textbf{76.9}/\textbf{76.9} & \textbf{83.3}/81.7 \\
\texttt{output} & 67.5/\textbf{67.9} & \textbf{73.6}/70.4 & n/a\,/\,75.9 & n/a\,/\,66.1 \\
\bottomrule
\end{tabular}
\end{table}

\textbf{GEMM results.} The agent-forged CuTeDSL kernels remain close to cuBLAS
on most target shapes and surpass it on several forward and backward passes.
The strongest results occur for the attention-output and FFN projections.
For example, the forged attention-output backward kernel reaches $73.0\%$
MFU, compared with $71.5\%$ for cuBLAS.

\textbf{Attention results.} The forged FlashAttention forward trails cuDNN,
FlashAttention-3, and FlashAttention-4 but leads Transformer Engine, while the
forged backward leads every baseline, including FlashAttention-3 and
FlashAttention-4. This backward result is especially significant because
attention backward carries twice the FLOPs of forward; Appendix~\ref{app:anatomy}
traces the full forging trajectory of this backward kernel, from the naive
atomic-accumulate baseline to the final warped-specialized, layout-refactored
implementation. The attention contribution
therefore comes mainly from the backward kernel rather than from a uniform
improvement in both directions.

\textbf{End-to-end result.} After the selected GEMM and FlashAttention kernels
are integrated into the same correctness-forged framework, the complete
MiniCPM4 0.5B engine reaches $44.13\%$ MFU, compared with $41.66\%$ before
operator-level forging. The operator stage therefore adds $2.47$ percentage
points of end-to-end MFU.

\subsubsection{Sources of the Speedup}\label{sec:engine-design}

The forged 0.5B engine's advantage over Megatron-LM~0.15 comes from three
sources. The first is a \emph{reproduced standard set}: optimizations that
exist in Megatron with the same boundaries, reproduced component for
component, including bucketed reduce-scatter overlapped with backward, ZeRO-1 optimizer
sharding, fused RMSNorm, RoPE, SwiGLU, and cross-entropy kernels, and a
direct Transformer-Engine attention path. The harness rediscovered them in the
first nine steps of the framework-level optimization trajectory
(Table~\ref{tab:framework-opt}, Appendix~\ref{app:framework-opt}).

The second is \emph{vertical integration}: components that Megatron also has
but keeps as separate stages, which the forged engine merges into a single
execution path for the fixed scenario. Megatron's optimizer tail runs
gradient clipping, the Adam update, and the FP32-to-BF16 parameter copy as
three passes over every parameter; the forged engine fuses them into one
Triton kernel with a single pass. Likewise, rather than graphing individual
transformer layers, the forged engine captures a whole microbatch's forward,
loss, and backward as one CUDA Graph replay.

The third is \emph{deep customization}: scenario-specific points with no
counterpart in Megatron, where the forged engine rewrites the execution path
around the fixed model, shapes, and parallel layout. Optimizer scalars and
the clipping coefficient stay on the device, so the step never synchronizes
with the host; the loss all-reduce is moved off the critical path; and the
read-only and layout-specific paths are specialized to the fixed shapes.

Each mechanism's source locations, configuration gates, and the per-step
MFU delta it contributes are detailed in Appendix~\ref{app:engine-design}.

\subsection{End-to-End Correctness Validation}\label{sec:validation}

\begin{figure}[t]
\centering
\includegraphics[width=\linewidth]{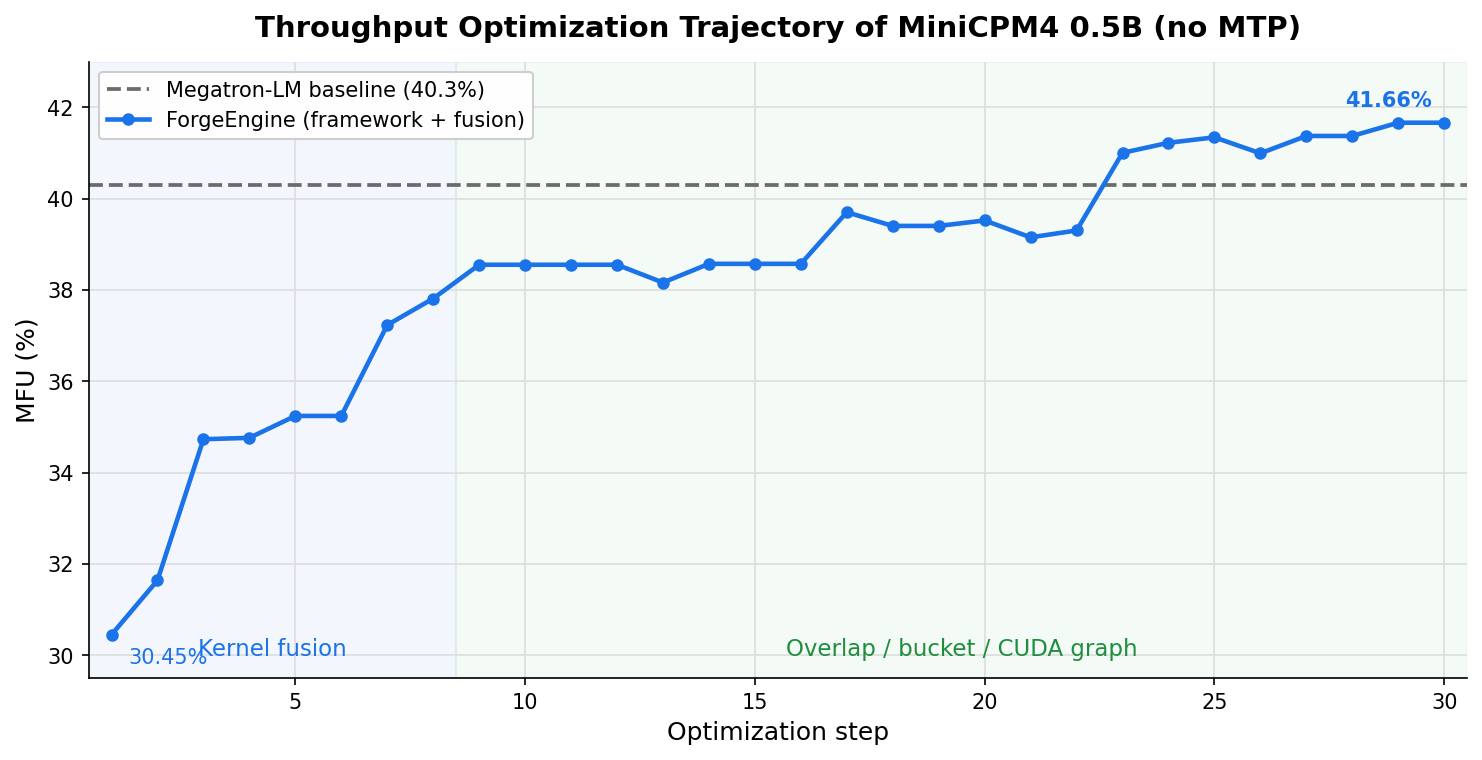}
\caption{Framework-level forging trajectory of the MiniCPM4 0.5B engine on
$16\times$H100. Each point is an accepted optimization step; MFU rises from
$30.45\%$ to $41.66\%$ and crosses the Megatron-LM baseline of $40.3\%$
(dashed) after the kernel-fusion steps give way to overlap, bucketing, and
CUDA-graph optimizations. This is the engine used for the MiniCPM4 0.5B
correctness runs below.}
\label{fig:mfu-curve}
\end{figure}

We evaluate ForgeTrain's correctness end to end by using the forged engines
for real production training and comparing their training trajectories,
resulting checkpoints, and downstream model quality with those of the trusted
reference implementations described in Section~\ref{sec:setup}. The engines
under test are the framework-forged engines running at full production
throughput: the MiniCPM4 0.5B engine trains at $41.66\%$ MFU on
$16\times$H100 against $40.3\%$ for Megatron-LM
(Figure~\ref{fig:mfu-curve} traces how forging raised it from $30.45\%$ to
this level), and the MiniCPM5 1B and MiniCPM5 130M engines train on Ascend
910 at $37.0\%$ MFU on $32$ devices (DP$=32$) against $33.6\%$ for MindSpeed
and at $24.9\%$ MFU on $2$ devices (DP$=2$) against $23.0\%$ for MindSpeed.
In wall-clock terms, the MiniCPM4 0.5B engine first passed Megatron-LM after
roughly ten hours of forging and reached this level after two to three
further days.

\subsubsection{Training-Trajectory Agreement}\label{sec:e2e}

\begin{figure}[t]
\centering
\includegraphics[width=\linewidth]{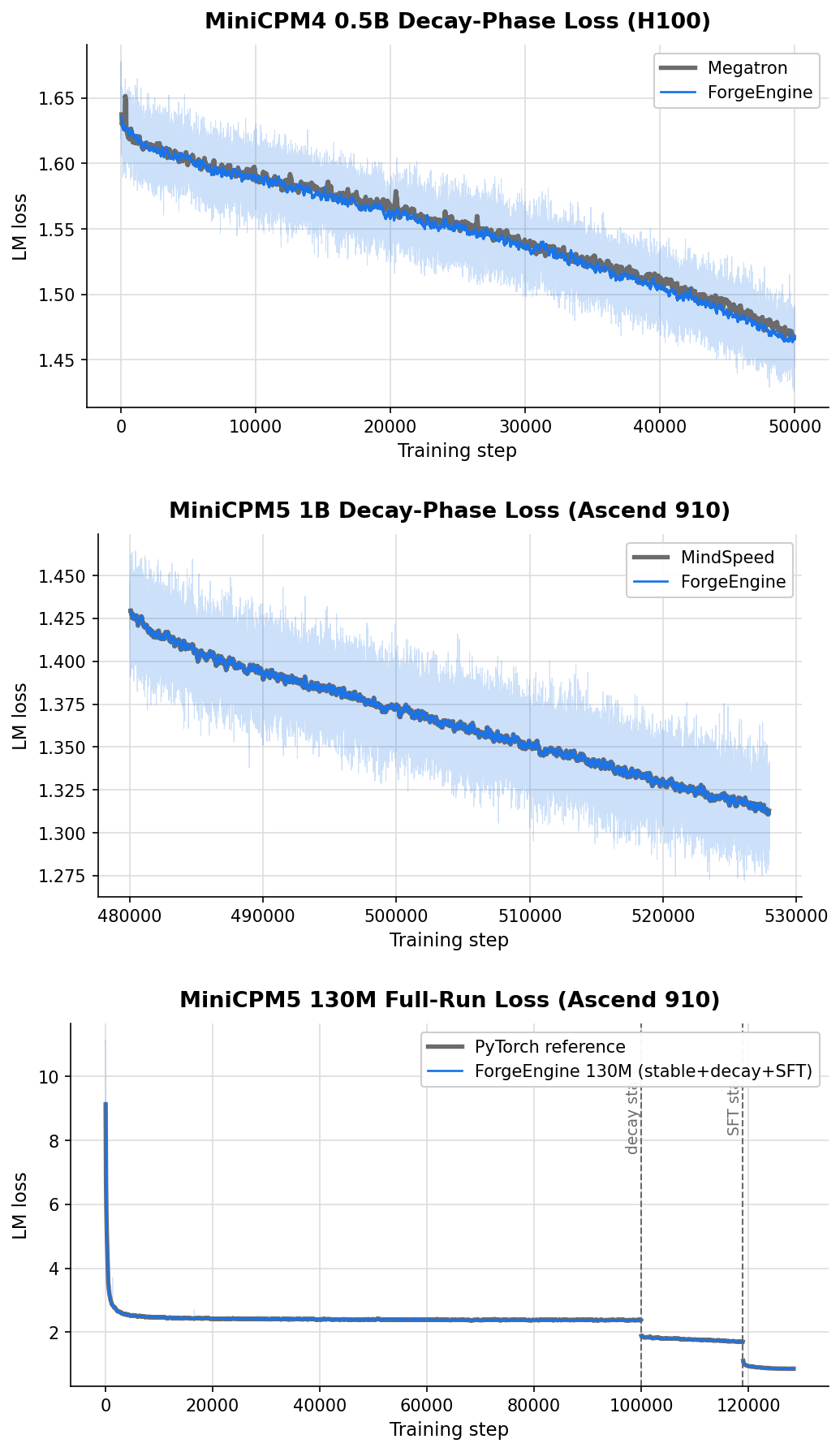}
\caption{Loss trajectories for the three long-horizon comparisons: MiniCPM4
0.5B decay on H100, MiniCPM5 1B decay on Ascend 910, and the complete
MiniCPM5 130M stable--decay--SFT run on Ascend 910.}
\label{fig:loss-validation}
\end{figure}

We compare loss trajectories under matched checkpoints, data, and schedules.
Figure~\ref{fig:loss-validation} summarizes the three runs. For
MiniCPM4 0.5B, we resume the decay phase from a shared stable-phase checkpoint
and compare with Megatron-LM. For MiniCPM5 1B, we resume from the same
480k-step checkpoint and run the same 48{,}000-step decay schedule on Ascend;
the final losses are $1.3227$ and $1.3226$. For MiniCPM5 130M, the forged
engine runs the complete stable--decay--SFT pipeline on Ascend, and its loss
trajectory is compared with a PyTorch reference over the same schedule. The
final SFT losses are $0.845$ and $0.861$.

\subsubsection{Downstream Model Quality}\label{sec:downstream}

For MiniCPM4 0.5B, we fine-tune the decay-phase checkpoints from both runs
with the same Megatron-LM SFT procedure and evaluate the resulting models on
six benchmarks. This keeps SFT fixed and isolates the effect of the
pretraining engine. The model trained with the forged engine matches or
exceeds the Megatron baseline on five of the six benchmarks.

\begin{table}[t]
\centering
\small
\caption{Downstream evaluation of MiniCPM4 0.5B on H100 after the same
decay schedule with Megatron-LM and with the forged engine, followed by
identical Megatron-LM SFT.}
\label{tab:downstream}
\begin{tabular}{@{}lccc@{}}
\toprule
\textbf{Benchmark} & \textbf{Megatron} & \textbf{ForgeTrain} & \textbf{$\Delta$} \\
\midrule
CMMLU              & 65.23 & 65.69 & $+0.46$ \\
MBPP (sanitized)   & 57.98 & 57.98 & $0.00$  \\
GSM8K              & 49.96 & 50.95 & $+0.99$ \\
MATH (PRM800K-500) & 31.00 & 32.20 & $+1.20$ \\
IFEval             & 50.65 & 50.28 & $-0.37$ \\
C-Eval (CoT)       & 64.16 & 65.94 & $+1.78$ \\
\midrule
\textbf{Average}   & \textbf{53.16} & \textbf{53.84} & \textbf{$+0.68$} \\
\bottomrule
\end{tabular}
\end{table}

For MiniCPM5 1B on Ascend, we evaluate the checkpoints produced by the two
decay runs in Section~\ref{sec:e2e} on eight benchmarks, covering Chinese and
English knowledge (C-Eval, MMLU-Redux), code (HumanEval, MBPP), mathematical
reasoning (GSM8K, MATH-500), general reasoning (BBH), and instruction
following (IFEval). Table~\ref{tab:downstream-ascend} reports the results.
The model trained with the forged engine exceeds the MindSpeed baseline on
four benchmarks and trails it on the other four.

\begin{table}[t]
\centering
\small
\caption{Downstream evaluation of MiniCPM5 1B on Ascend 910 after the
same decay schedule with MindSpeed and with the forged engine.}
\label{tab:downstream-ascend}
\begin{tabular}{@{}lccc@{}}
\toprule
\textbf{Benchmark} & \textbf{MindSpeed} & \textbf{ForgeTrain} & \textbf{$\Delta$} \\
\midrule
C-Eval             & 37.10 & 35.53 & $-1.57$ \\
MMLU-Redux         & 42.90 & 42.41 & $-0.49$ \\
HumanEval          & 36.59 & 37.20 & $+0.61$ \\
MBPP (sanitized)   & 44.75 & 48.25 & $+3.50$ \\
GSM8K              & 58.91 & 54.59 & $-4.32$ \\
MATH-500           & 30.20 & 31.40 & $+1.20$ \\
BBH                & 26.72 & 22.01 & $-4.71$ \\
IFEval             & 34.01 & 34.38 & $+0.37$ \\
\midrule
\textbf{Average}   & \textbf{38.90} & \textbf{38.22} & \textbf{$-0.68$} \\
\bottomrule
\end{tabular}
\end{table}

For MiniCPM5 130M on Ascend, both the forged engine and the PyTorch
reference run the complete stable--decay--SFT pipeline in
Section~\ref{sec:e2e}, so this comparison also covers the SFT stage. We
evaluate the final checkpoints on five English benchmarks spanning general
reasoning (BBH), mathematical reasoning (GSM8K, MATH), and code (HumanEval,
MBPP). Table~\ref{tab:downstream-ascend-130m} reports the results. The
model trained with the forged engine exceeds the reference on four of the
five benchmarks. Across
the three comparisons, the average difference stays within one point and
changes sign, so the forged engines are interchangeable with their references
at the level of downstream quality on both hardware ecosystems.

\begin{table}[t]
\centering
\small
\caption{Downstream evaluation of MiniCPM5 130M on Ascend 910 after the
complete stable--decay--SFT pipeline with the PyTorch (\texttt{torch\_npu})
reference and with the forged engine.}
\label{tab:downstream-ascend-130m}
\begin{tabular}{@{}lccc@{}}
\toprule
\textbf{Benchmark} & \textbf{PyTorch} & \textbf{ForgeTrain} & \textbf{$\Delta$} \\
\midrule
BBH                & 32.93 & 32.08 & $-0.85$ \\
GSM8K              & 17.29 & 18.42 & $+1.13$ \\
MATH               &  9.80 & 10.20 & $+0.40$ \\
HumanEval          & 20.12 & 21.34 & $+1.22$ \\
MBPP               & 34.63 & 37.35 & $+2.72$ \\
\midrule
\textbf{Average}   & \textbf{22.95} & \textbf{23.88} & \textbf{$+0.93$} \\
\bottomrule
\end{tabular}
\end{table}

\subsection{Ablation}\label{sec:ablation}

We examine two reduced settings with Claude Opus~5 (1M-token context
window) on H100. The first
removes the Harness for the MiniCPM5 16B A3B task in Section~\ref{sec:throughput}.
The second retains the Milestones but removes the quality constraints for
MiniCPM4 0.5B. Each is compared with the complete Harness on its corresponding
model; the two ablations address different failure modes.

\textbf{Without the Harness.} During the MiniCPM5 16B A3B implementation, the agent
reduces optimizer precision from FP32 to BF16 to accelerate training. This
changes the numerical computation of the optimizer, so throughput alone
cannot establish that the resulting engine preserves training quality.
The run therefore provides no verified performance gain at training-quality
parity.

\begin{figure}[t]
\centering
\includegraphics[width=\linewidth]{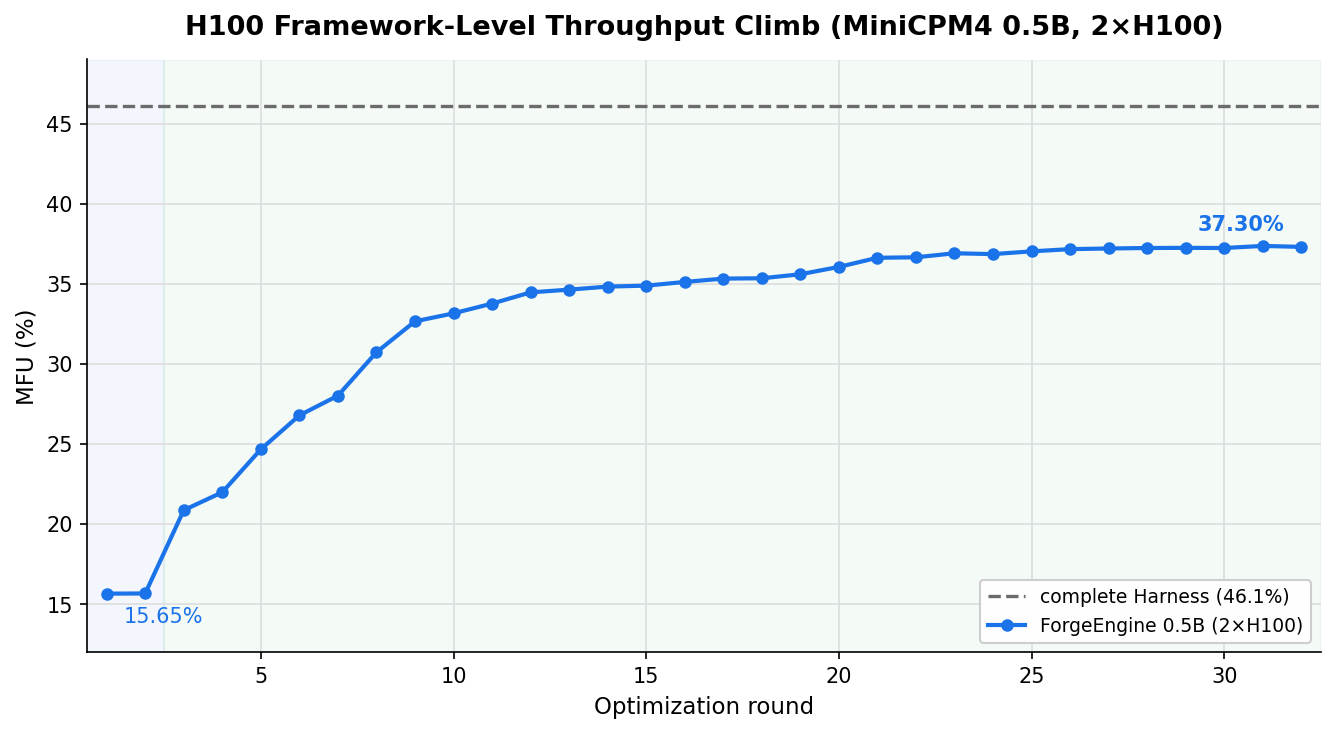}
\caption{MiniCPM4 0.5B forging without quality constraints. The dashed line is
the complete-Harness engine of Section~\ref{sec:throughput}, at $46.1\%$ on
the same $2\times$H100 setting.}
\Description{MFU rises during early rounds and plateaus near 37 percent by
round 32, far below the dashed 46.1 percent line.}
\label{fig:ablation-milestones}
\end{figure}

\textbf{Without quality constraints.} On MiniCPM4 0.5B, MFU remains near
$37\%$ within 33 forging rounds, far below the $46.1\%$ that the complete
Harness reaches for the same setting (Table~\ref{tab:mfu-checkpoints}).
Figure~\ref{fig:ablation-milestones}
shows the available trajectory through round~32. The agent's reasoning
repeatedly treats Bit-for-Bit agreement as a restriction on optimization.
In round~18, it rules out changing deterministic attention backward because
of its effect on numerical agreement. In round~32, it again treats attention
and GEMM implementations as fixed and considers changes to the reduction
order too risky. The search consequently concentrates on smaller changes
that preserve the existing arithmetic.

Together, these observations illustrate why the Harness needs an explicit
quality criterion. Without one, the agent may either change optimizer
precision without establishing training-quality parity, or continue to
require exact numerical agreement even during performance optimization.
ForgeTrain specifies the transition from Bit-for-Bit agreement to
training-quality parity, giving the agent a criterion for evaluating changes
to the reference computation.

%% file: sections/related.tex
\section{Related Work}\label{sec:related}


\subsection{Large-Scale Training Frameworks}\label{sec:related-frameworks}

ForgeTrain forges against a mature body of human-engineered training
systems. Megatron-LM~\citep{shoeybi2019megatron} popularized tensor and
pipeline model parallelism for transformers; DeepSpeed's ZeRO partitions
optimizer states, gradients, and parameters across data-parallel ranks to
fit trillion-parameter models~\citep{rajbhandari2020zero}; GPipe
introduced micro-batch pipeline parallelism with synchronous
updates~\citep{huang2019gpipe}; and Alpa automatically searches a
hierarchical space of inter- and intra-operator parallel
plans~\citep{zheng2022alpa}. These are the canonical incumbents of the
one-codebase paradigm: a single general framework whose abstraction layers
and configuration knobs are built to cover every model, scale, and hardware
topology. Even Alpa's automation searches over a fixed framework's plan
space rather than regenerating the framework itself. ForgeTrain inverts
this: instead of one framework amortized across all scenarios, it forges a
per-scenario ForgeEngine from scratch.

\subsection{AI-Generated System Software}\label{sec:related-systems}

The closest neighbors to ForgeTrain are whole-system software artifacts
written end-to-end by AI. VibeTensor~\citep{vibetensor2026} is presented
as the first fully AI-generated deep learning system, a PyTorch-style
eager runtime assembled by agents. It is, however, released for
agentic-systems research only: it runs $1.7$--$6.2{\times}$ slower than
PyTorch, and its authors describe a ``Frankenstein composition effect''
in which locally correct components compose into a globally suboptimal
whole. The shortfall is not implementation skill but the absence of a
truth source that forces global behavior to
match a production baseline.

Claude's C Compiler~\citep{anthropic_ccc2026} is methodologically the
closest relative: roughly one hundred thousand lines of Rust written by
sixteen cooperating agents, compiling the Linux 6.9 kernel, with GCC used
as a differential-testing oracle. Promoting a reference implementation to a
golden reference is precisely the discipline ForgeTrain formalizes in its Bit-for-Bit
stage. Yet the
compiler falls back to GCC for assembly and linking, is explicitly
disclaimed as not production-ready, and targets the compiler domain
rather than training infrastructure. ForgeTrain differs on all three axes
these systems leave open: it is whole-system, production-grade, and
matches and surpasses its reference on the reference's own
benchmarks.

\subsection{AI-Generated Kernels and Operators}\label{sec:related-kernels}

A productive line of work forges performance-critical code at node
granularity. AlphaEvolve~\citep{alphaevolve2025} couples an evolutionary
search loop with automated evaluation and has discovered scheduling
heuristics and matrix-multiplication kernels deployed in production;
KernelEvolve~\citep{kernelevolve2026} evolves production-grade kernels
under throughput targets; and Sakana's CUDA
Engineer~\citep{sakana_cuda2025} synthesizes thousands of test-verified
CUDA kernels. The combination of iterative or evolutionary search,
test-driven verification, and profiling feedback used in these systems is
precisely the
Harness-driven loop, and these systems show it is already production-ready
at the level of a single kernel or operator.

The distinction from ForgeTrain is granularity. Each of these efforts
forges one self-contained node with a clear local objective, where
correctness and speed can be measured in isolation. A training framework
is an integrated system whose behavior is
correct only globally, across an entire optimization trajectory. Forging
at that granularity is what the match-then-surpass discipline of
\S\ref{sec:method} makes possible, and it is the gap these node-level
efforts leave open.

%% file: sections/conclusion.tex
\section{Discussion and Conclusion}\label{sec:conclusion}

Across seven model--hardware settings on two hardware ecosystems, every
forged ForgeEngine surpasses its golden reference by $4.7$--$33.2\%$ MFU
at matched training quality. MiniCPM4 0.5B weights trained on a
ForgeEngine match the Megatron-LM-trained baseline on downstream
evaluation, and on Ascend a forged engine carried a 130M model through a
complete stable--decay--SFT production run. ForgeTrain thus shows that a
production-grade training framework can be forged end-to-end by an
autonomous agent from an empty repository. To our knowledge, it is the
first such framework forged by AI to match and surpass its human
reference.

ForgeTrain is thoroughly validated at smaller model
scales, but forging for larger models remains to be demonstrated
experimentally, and the long-term maintenance and evolution of a forged
engine are untested.

ForgeTrain instantiates Forge Engineering: building dedicated
systems software from scratch for each scenario and iteratively optimizing
it toward peak performance under correctness and usability constraints.
A reusable Harness carries construction knowledge and executable
evaluations across scenarios, while each implementation is tailored to
its workload, hardware, and operational requirements. As coding agents
improve, Forge Engineering could make dedicated systems a practical
alternative to general-purpose implementations across a wider range of
domains.

%% file: sections/appendix.tex
\section{Framework-Level Optimization Trajectory}\label{app:framework-opt}

Table~\ref{tab:framework-opt} lists, in order of acceptance, the framework-level
optimizations the agent loop applied to the MiniCPM4 0.5B ForgeEngine during
Stage~C, together with the measured MFU after each step
(16$\times$H100, DP only). This is the per-step detail behind the trajectory of
Table~\ref{tab:framework-opt}: the first eight steps form the kernel-fusion regime, and
the remainder the overlap/bucket/CUDA-graph regime. Two candidates ($\dagger$)
were measured but rejected by the Stage-C gate and reverted, so they do not
carry into the final engine.

\begin{table*}[t]
\centering
\small
\caption{Framework-level optimizations applied to MiniCPM4 0.5B during Stage~C,
in acceptance order. MFU is measured on 16$\times$H100 (data-parallel).
$\dagger$~marks candidates that were reverted after the gate rejected them.}
\label{tab:framework-opt}
\begin{tabular}{@{}rll r@{}}
\toprule
\textbf{\#} & \textbf{Phase} & \textbf{Optimization} & \textbf{MFU (\%)} \\
\midrule
1  & Kernel fusion & Stage-B baseline (no fusion)            & 30.45 \\
2  & Kernel fusion & Fused cross-entropy + SwiGLU            & 31.64 \\
3  & Kernel fusion & Cross-entropy \texttt{NaN} + embedding-backward fix & 34.73 \\
4  & Kernel fusion & Direct Transformer-Engine attention     & 34.76 \\
5  & Kernel fusion & Fused residual-add + RMSNorm            & 35.24 \\
6  & Kernel fusion & Fused gradient-clip + Adam + param-sync & 35.24 \\
7  & Kernel fusion & \texttt{as\_strided} cross-entropy + FFN fusion & 37.23 \\
8  & Kernel fusion & Cross-entropy prefill + fused RMSNorm backward & 37.81 \\
\midrule
9  & Overlap/graph & Gradient bucketing (100\,MB)            & 38.55 \\
10 & Overlap/graph & Per-bucket all-gather + optimizer overlap & 38.55 \\
11 & Overlap/graph & All-gather--optimizer overlap           & 38.55 \\
12 & Overlap/graph & Wgrad-overlap by default                & 38.55 \\
13 & Overlap/graph & NCCL tuning + async loss all-reduce     & 38.16 \\
14 & Overlap/graph & Production-default verification          & 38.57 \\
15 & Overlap/graph & Batch-flush write-grad restructure      & 38.57 \\
16 & Overlap/graph & Batch-flush gate                        & 38.57 \\
17 & Overlap/graph & ZeRO-1 sharded optimizer                & 39.70 \\
18 & Overlap/graph & Sharded optimizer by default            & 39.40 \\
19 & Overlap/graph & Routing / baseline fixes                & 39.40 \\
20 & Overlap/graph & Fused norm-backward + multi-tensor Adam & 39.52 \\
21 & Overlap/graph & Fused-operator micro-optimizations      & 39.15 \\
22 & Overlap/graph & RoPE strided-K (drop K copy)            & 39.30 \\
23 & Overlap/graph & Document-aware fused RoPE               & 41.00 \\
24 & Overlap/graph & Wgrad bucket 200\,MB                    & 41.22 \\
25 & Overlap/graph & RoPE backward in-place$^{\dagger}$      & 41.34 \\
26 & Overlap/graph & Forward-only CUDA Graph                 & 40.99 \\
27 & Overlap/graph & Full-step CUDA Graph (phase A)          & 41.37 \\
28 & Overlap/graph & Full-step CUDA Graph (phase B/C)        & 41.37 \\
29 & Overlap/graph & Wgrad bucket 400\,MB (single bucket)    & 41.66 \\
30 & Overlap/graph & Cross-step pipelining$^{\dagger}$       & 41.66 \\
\bottomrule
\end{tabular}
\end{table*}

\begin{figure*}[t]
\centering
\begin{subfigure}[b]{0.19\textwidth}
\includegraphics[width=\linewidth]{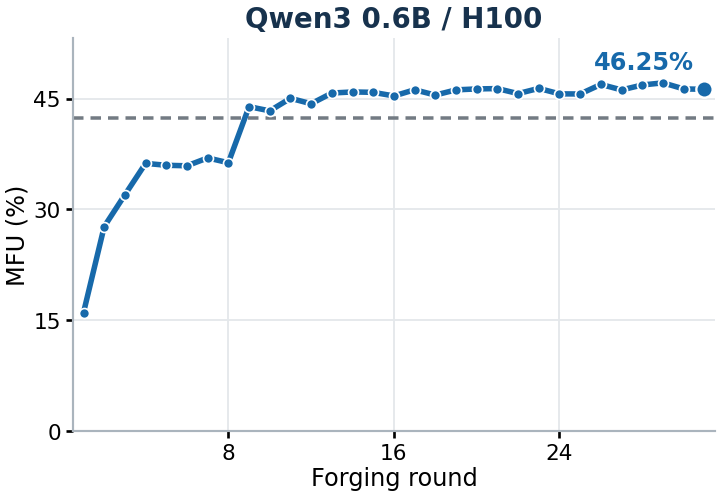}
\caption{Qwen3 0.6B}
\label{fig:mfu-qwen3}
\end{subfigure}\hfill
\begin{subfigure}[b]{0.19\textwidth}
\includegraphics[width=\linewidth]{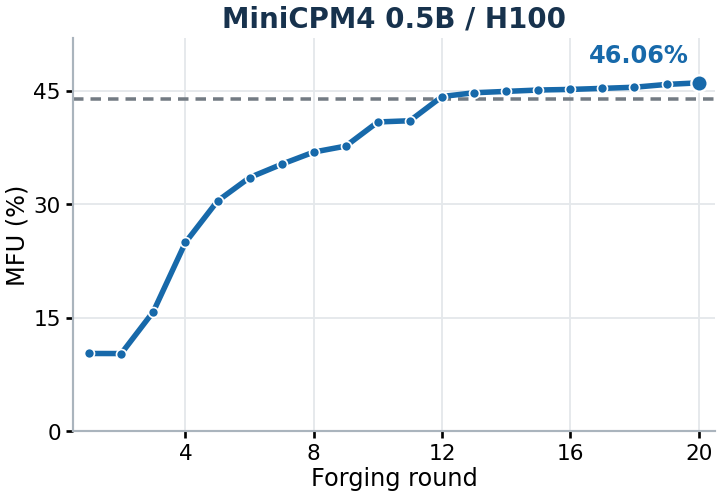}
\caption{MiniCPM4 0.5B}
\label{fig:mfu-05b}
\end{subfigure}\hfill
\begin{subfigure}[b]{0.19\textwidth}
\includegraphics[width=\linewidth]{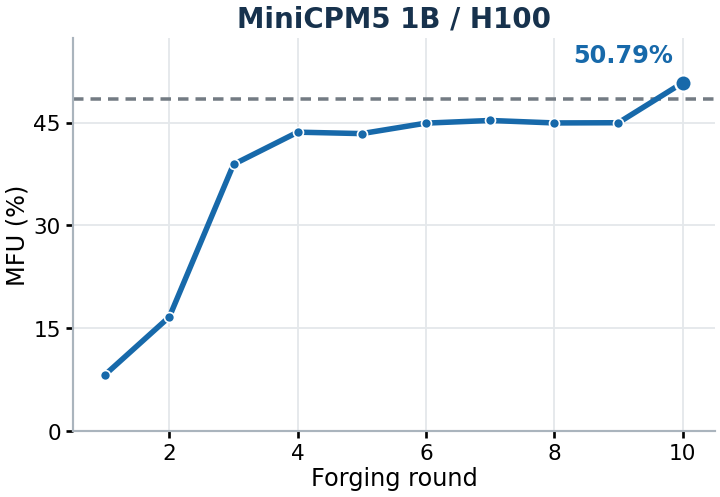}
\caption{MiniCPM5 1B}
\label{fig:mfu-1b}
\end{subfigure}\hfill
\begin{subfigure}[b]{0.19\textwidth}
\includegraphics[width=\linewidth]{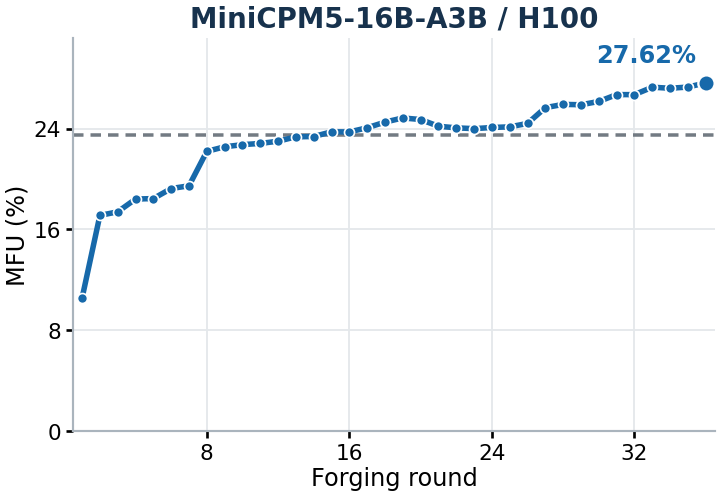}
\caption{MiniCPM5 16B A3B}
\label{fig:mfu-moe}
\end{subfigure}\hfill
\begin{subfigure}[b]{0.19\textwidth}
\includegraphics[width=\linewidth]{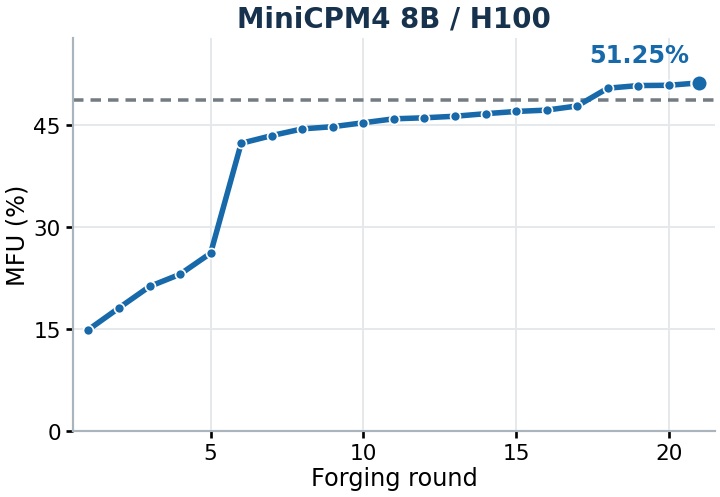}
\caption{MiniCPM4 8B}
\label{fig:mfu-8b}
\end{subfigure}
\caption{Per-setting MFU trajectories of the five H100 settings from
Table~\ref{tab:mfu-checkpoints}, each normalized to its own Megatron-LM
baseline (dashed). The x-axis is the setting's $k$-th recorded forging round.}
\label{fig:mfu-panels-h100}
\end{figure*}

\section{Model Architectures}\label{app:models}

The H100 evaluation scenarios use three released model families without
architectural modification: MiniCPM4 0.5B and 8B
\footnote{\url{https://www.modelscope.cn/models/OpenBMB/MiniCPM4-0.5B} and
\url{https://www.modelscope.cn/models/OpenBMB/MiniCPM4-8B}.}, Qwen3 0.6B
\footnote{\url{https://huggingface.co/Qwen/Qwen3-0.6B}.}, and MiniCPM5 1B and
16B-A3B. All are decoder-only
transformers with grouped-query attention, SwiGLU feed-forward blocks, and
pre-normalization RMSNorm; MiniCPM4 uses LongRoPE positional encoding and the
MiniCPM $\mu$P-style scaling (\texttt{scale\_emb}, \texttt{dim\_model\_base},
\texttt{scale\_depth}), Qwen3 uses standard parameterization with RoPE
($\theta{=}10^6$) and per-head QK-Norm, and MiniCPM5 is a dense or sparse
(MoE) model with RoPE and $\mu$P scaling. Tables~\ref{tab:arch} and
\ref{tab:h100-arch} give the configurations that fix the FLOP count and the
shapes the forged operators target; the experiments train at a sequence length
of $4096$ (\S\ref{sec:setup}), within the context each architecture supports.

\begin{table*}[t]
\centering
\small
\caption{MiniCPM4 architecture configuration for the two evaluation
scenarios.$^{a}$}
\label{tab:arch}
\begin{tabular}{@{}lcc@{}}
\toprule
\textbf{Hyperparameter} & \textbf{MiniCPM4 0.5B} & \textbf{MiniCPM4 8B} \\
\midrule
Transformer layers              & 24    & 32    \\
Hidden size                     & 1024  & 4096  \\
FFN intermediate size           & 4096  & 16384 \\
Attention heads                 & 16    & 32    \\
Key/value heads (GQA)           & 2     & 2     \\
Head dimension                  & 64    & 128   \\
Vocabulary size                 & 73448 & 73448 \\
Activation                      & SwiGLU (SiLU) & SwiGLU (SiLU) \\
Normalization                   & RMSNorm ($\epsilon{=}10^{-5}$) & RMSNorm ($\epsilon{=}10^{-6}$) \\
Position encoding               & LongRoPE & LongRoPE \\
Max context length              & 32768 & 32768 \\
Tied input/output embeddings    & Yes   & No    \\
$\mu$P (\texttt{scale\_emb}, \texttt{dim\_model\_base}, \texttt{scale\_depth}) & $12,256,1.4$ & $12,256,1.4$ \\
MTP / Eagle head                & On (1 layer) & Off \\
Training precision              & bfloat16 & bfloat16 \\
\bottomrule
\end{tabular}
\vspace{2pt}
\par\small\textit{$^{a}$The 0.5B column carries a one-layer MTP (Eagle) head. The
framework-level setting of Section~\ref{sec:throughput} forges it, whereas the
operator-level forge of Section~\ref{sec:operators} and the correctness
experiment of Section~\ref{sec:validation} run the same model with this head
disabled (no-MTP). The 8B column trains without MTP.}
\end{table*}

\begin{table*}[t]
\centering
\small
\caption{H100 architecture configurations for the Qwen3 and MiniCPM5
scenarios. MiniCPM5 16B-A3B is a sparse MoE model (160 routed experts,
top-16, plus a 512-wide shared expert); its dense FFN width applies to
layer~0 only, and layers~1--27 use \texttt{moe\_ffn\_hidden\_size} instead.}
\label{tab:h100-arch}
\begin{tabular}{@{}lccc@{}}
\toprule
\textbf{Hyperparameter} & \textbf{Qwen3 0.6B} & \textbf{MiniCPM5 1B} & \textbf{MiniCPM5 16B-A3B} \\
\midrule
Transformer layers              & 28    & 24    & 28    \\
Hidden size                     & 1024  & 1536  & 2048  \\
FFN intermediate size           & 3072  & 4608  & 8192  \\
MoE FFN size (routed experts)   & ---   & ---   & 512   \\
Routed experts (top-$k$)        & ---   & ---   & 160 (16) \\
Attention heads                 & 16    & 16    & 32    \\
Key/value heads (GQA)           & 8     & 2     & 2     \\
Head dimension                  & 128   & 128   & 128   \\
Vocabulary size                 & 151936 & 130560 & 130560 \\
Activation                      & SwiGLU (SiLU) & SwiGLU (SiLU) & SwiGLU (SiLU) \\
Normalization                   & RMSNorm ($\epsilon{=}10^{-6}$) & RMSNorm ($\epsilon{=}10^{-6}$) & RMSNorm ($\epsilon{=}10^{-6}$) \\
Position encoding               & RoPE ($\theta{=}10^6$, QK-Norm) & RoPE & RoPE \\
Tied input/output embeddings    & Yes   & No    & No    \\
$\mu$P (\texttt{scale\_emb}, \texttt{dim\_model\_base}, \texttt{scale\_depth}) & ---   & $12,256,1.4$ & $12,256,1.4$ \\
MTP / Eagle head                & Off   & On (1 layer) & Off \\
Training precision              & bfloat16 & bfloat16 & bfloat16 \\
\bottomrule
\end{tabular}
\end{table*}

\section{Model-FLOPs Utilization}\label{app:flops}

We report MFU against an exact per-operator FLOP count rather than the usual
closed-form $6ND$ approximation, which omits attention and miscounts the
GQA projections and the vocabulary head. We count a multiply--accumulate as
two FLOPs and report per token; a quantity is the same for every token in the
batch, so the per-step numerator is this total times the number of tokens
processed. Let $h$ be the hidden size, $f$ the FFN intermediate size, $n_h$ and
$n_{kv}$ the query and key/value head counts, $d$ the head dimension
($h_q{=}n_h d$, $h_{kv}{=}n_{kv}d$), $V$ the vocabulary size, $L$ the layer
count, and $s$ the sequence length. The forward per-token FLOPs of each operator
class are
\begin{align*}
\text{Attn.\ projections (Q,K,V,O)} &: \; 2h\,(h_q + 2h_{kv}) + 2h_q h, \\
\text{FFN (gate, up, down; SwiGLU)} &: \; 2\,(2hf) + 2fh, \\
\text{Attention core (}QK^\top,\,AV\text{; causal)} &: \; 2 n_h s d, \\
\text{Output / cross-entropy head} &: \; 2hV.
\end{align*}
The attention core carries a factor $\tfrac12$ for the causal mask (folded into
the expression above), and grows with $s$ where the others do not. The backward
pass costs twice the forward for every GEMM (one matrix for the data gradient,
one for the weight gradient) and twice the forward for the attention core, so
each per-layer term is multiplied by three (one forward, two backward) and
summed over $L$ layers; the output head is counted once, also at the
$1{:}2$ forward-to-backward ratio. Table~\ref{tab:flops} evaluates this for the
two evaluation scenarios.

\begin{table*}[t]
\centering
\small
\caption{Per-token FLOP budget (forward${+}$backward) by operator class for the
two MiniCPM4 scenarios at sequence length $4096$, with each class's share of the
per-step total. The total is the numerator of the MFU computed in
\S\ref{sec:setup}.}
\label{tab:flops}
\begin{tabular}{@{}lcccc@{}}
\toprule
& \multicolumn{2}{c}{\textbf{0.5B}} & \multicolumn{2}{c}{\textbf{8B}} \\
\cmidrule(lr){2-3}\cmidrule(lr){4-5}
\textbf{Operator class} & \textbf{GFLOP/tok} & \textbf{Share} & \textbf{GFLOP/tok} & \textbf{Share} \\
\midrule
FFN GEMM (gate/up/down)           & $1.812$ & $56.5\%$ & $38.655$ & $76.5\%$ \\
Attention core (FlashAttention)   & $0.604$ & $18.8\%$ & $3.221$  & $6.4\%$ \\
Output / CE GEMM                  & $0.451$ & $14.1\%$ & $1.805$  & $3.6\%$ \\
Attn.\ projection GEMM (Q/K/V/O)  & $0.340$ & $10.6\%$ & $6.845$  & $13.5\%$ \\
\midrule
\textbf{Total}                    & $3.207$ & $100\%$  & $50.526$ & $100\%$ \\
\bottomrule
\end{tabular}
\end{table*}

Two properties of this count matter for the operator budget of
\S\ref{sec:operators}. First, the GEMMs and the attention core together are the
entire compute FLOP budget---normalization, RoPE, activations, and the optimizer
contribute negligible FLOPs and are memory-bound, which is why they are absent
from this table yet still consume measurable time (the elementwise bucket of
Table~\ref{tab:opbudget}). Second, the attention core is the one term that scales
with $s$: at the $4096$ sequence length it is $18.8\%$ of the 0.5B FLOP budget,
and two-thirds of that is backward, which is why the attention backward is the
single operator whose forging trajectory we trace in detail
(Appendix~\ref{app:anatomy}).

\section{Anatomy of a Forged Kernel}\label{app:anatomy}

\input{figures/fig_attn_bwd}

The attention backward (\S\ref{sec:operators}) is the clearest
illustration of how the loop forges a kernel, and its trajectory
(Figure~\ref{fig:attn-bwd}) separates two kinds of decision. We optimize it on an
H100~SXM at the engine's attention shape ($B{=}10$, $H{=}16$, $N{=}4096$,
$D{=}64$, FP16, GQA $8{:}1$, causal), organized---following the FA-3/FA-4
design---as a three-kernel pipeline (preprocess~$\rightarrow$~main~$\rightarrow$
postprocess). From a naive baseline that accumulates $\mathrm{d}Q$ through
\texttt{atomicAdd} ($5.19$\,ms, $16.5\%$ MFU), the loop first lays a
\emph{structural} foundation: a $128$-row $M$-tile (versus $64$), warp
specialization into one producer and two consumer warp-groups ($384$ threads, a
$24/240$-register split), double-buffering on every pipeline stage, and an
asynchronous five-GEMM issue schedule. These choices are mutually dependent and
do not pay off in isolation---while they are being put in place latency
regresses to $6.3$--$6.9$\,ms, because the $\mathrm{d}Q$-accumulator
data-flow is still the bottleneck. The decisive step is a data-layout refactor
that replaces the WGMMA-fragment-aware $\mathrm{d}Q$ accumulator with a flat
global buffer and a flat thread-value register-to-shared copy, and sources the
$\mathrm{d}K/\mathrm{d}V$ GEMM operands directly from registers; this cuts
shared-memory bank conflicts from $73$\,M to $8.2$\,M and brings the kernel to
$2.22$\,ms ($38.6\%$ MFU). Three independent \emph{tactical} optimizations then
close the gap: $128$-bit vectorized preprocess/postprocess copies
($2.03$\,ms), skipping the causal mask on all but the diagonal block---which
halves arithmetic-pipe instructions ($-54\%$)---($1.97$\,ms), and removing a
stray \texttt{cudaStreamSynchronize} that had disabled programmatic dependent
launch, restoring cross-kernel overlap ($1.90$\,ms, $45.7\%$ MFU). The result is
$5.6\%$ faster than FlashAttention-4 ($2.007$\,ms, $43.3\%$ MFU) and $31\%$
faster than cuDNN ($2.483$\,ms, $35.0\%$ MFU); note that the margin over FA-4
runs the other way in the per-kernel NCU timings, which are \emph{slower} than
FA-4's ($2072$ versus $1990$\,\textmu s). The entire end-to-end margin therefore
comes from cross-kernel launch overlap: programmatic dependent launch hides
$0.17$\,ms across our four kernels, against no measurable overlap for FA-4's
five. The methodological point generalizes beyond this
one kernel: structural decisions must be fixed jointly and up front, even through
a transient regression, whereas tactical optimizations compose independently on
top of a correct structure.

\section{Optimization Catalog}\label{app:opt-catalog}

This appendix details the optimizations summarized in
\S\ref{sec:throughput} and \S\ref{sec:operators}.

\textbf{Framework level, second class (goals the reference shares, boundaries
redrawn on measured returns).}
\begin{itemize}[leftmargin=*,topsep=2pt,itemsep=3pt]
  \item \textbf{Whole-forward graph capture.} The roughly three hundred kernel
    launches from the embedding to the output projection are captured into a
    single CUDA graph replayed once per step, with the backward pass,
    communication, and the optimizer left outside. Megatron-LM offers a wider
    graph spectrum---per-layer graphs, a full-iteration graph that includes
    gradient reduction, and a separately graphed optimizer step; the agent's
    narrower boundary is an adjudicated outcome: the wider candidates
    (forward-plus-backward graphs, collectives and optimizer in-graph) were
    each tried in earnest and rejected by the gate for negative measured
    returns or an unsupporting runtime stack.
  \item \textbf{Fused optimizer step.} Gradient clipping, the AdamW update, and
    the master-to-\texttt{bf16} write-back are fused into a single kernel
    (\texttt{fused\_clip\_adam\_sync}), where Megatron-LM keeps the three as
    separate steps.
  \item \textbf{Optimizer--gather pipeline.} On the 8B engine, the parameter
    all-gather is interleaved with the optimizer step bucket by bucket: the
    moment a bucket's Adam update completes, the all-gather of its BF16
    parameters is issued on the communication stream, ordered by forward
    consumption so it overlaps the next step's forward, which waits only at
    bucket boundaries on the matching events. Megatron-LM's
    \texttt{overlap\_param\_gather} likewise overlaps the all-gather with the
    next forward, but dispatches it lazily, bucket by bucket, from forward
    pre-hooks; the agent moved the dispatch into the optimizer step itself,
    interleaved with per-bucket Adam as a single pipeline.
\end{itemize}

\textbf{Framework level, third class (no Megatron-LM counterpart).}
\begin{itemize}[leftmargin=*,topsep=2pt,itemsep=3pt]
  \item \textbf{Asynchronous loss all-reduce.} The scalar loss all-reduce is
    issued on the communication stream and overlapped with the optimizer step,
    hiding it entirely in measurement; Megatron-LM leaves this reduction
    synchronously on the critical path.
  \item \textbf{Device-side gradient norm and clipping.} The gradient norm,
    clip coefficient, and scaling are computed entirely on device---l2norm,
    all-reduce, square root, clamp, and scale in one uninterrupted
    pipeline---where Megatron-LM reads the norm back to the host
    (\texttt{.item()}) between norm and clip, placing a CPU--GPU
    synchronization on the hot path.
  \item \textbf{Zero-copy gradient views.} Having established that its fused
    Adam kernel only reads gradients, the agent had the all-gather return
    views rather than copies, eliminating roughly one hundred and fifty clones
    and about 2\,GB of peak memory per step.
  \item \textbf{Per-shape operator selection.} A dispatcher benchmarks candidate
    implementations across CuTeDSL, cuBLAS, Triton, and Transformer Engine and
    selects, per operator shape, the kernel of highest model-FLOPs utilization
    (MFU); Megatron-LM relies on a fixed backend with a small set of flags and
    performs no such automatic selection.
  \item \textbf{High-priority auxiliary streams.} The streams carrying wgrad
    and gradient reduction are set to high CUDA priority, letting them land
    ahead of the main stream's large GEMMs at SM scheduling boundaries and
    shortening the exposed communication and wgrad tail at the end of each
    step; Megatron-LM runs all its streams at default priority.
\end{itemize}

\textbf{Kernel level, the four recurring families.}
\begin{itemize}[leftmargin=*,topsep=2pt,itemsep=3pt]
  \item \textbf{Warp specialization and software pipelining.} Threads within a
    block are split into producer roles that move data and consumer roles that
    compute, with every pipeline stage multiply-buffered and the constituent
    matrix multiplies issued asynchronously, so memory movement and computation
    overlap across stages rather than serializing. A general-purpose library
    fixes one thread-to-work mapping for a whole shape class; forging it for the
    engine's exact dimensions recovers the occupancy that mapping leaves on the
    table.
  \item \textbf{Memory-layout specialization.} On-chip data is laid out for the
    operator's exact shape---flattening accumulator buffers to remove
    shared-memory bank conflicts, vectorizing the load/store paths, and sourcing
    matrix-multiply operands directly from registers rather than staging them
    through shared memory. These are the layout choices a vendor kernel cannot
    hard-code without knowing the shape, and they convert otherwise-stalled
    memory pipes into useful throughput.
  \item \textbf{Backward-pass fusion.} An operator's backward is computed by two
    matrix multiplies---one for the data gradient, one for the weight
    gradient---and, under sharding, a gradient reduction; the agent folds all
    three into a single kernel, sparing the launch overhead and intermediate
    write-back that a vendor library's separate \texttt{dgrad}, \texttt{wgrad},
    and reduce kernels pay.
  \item \textbf{Cross-kernel scheduling.} Adjacent kernels are overlapped across
    their launch boundary through programmatic dependent launch, and computation
    that the problem structure renders redundant is pruned outright (for a causal
    operator, evaluating the mask only on the diagonal block). Both recover time
    that lives between or inside kernels, where a single-kernel benchmark cannot
    reach.
\end{itemize}
\section{Why the Engine Is Faster, Mechanism by Mechanism}\label{app:engine-design}

\S\ref{sec:engine-design} traces the forged 0.5B engine's margin over
Megatron-LM~0.15 to three sources. This appendix expands each source with its
source location and key code. Citations are file:line into the exported
ForgeEngine package; snippets are trimmed of unrelated lines only.

\textbf{Reproduced standard set.} The optimizations Megatron-LM already has,
reproduced component for component with the same boundaries
(\texttt{nccl.py:152,654} --- bucketed reduce-scatter overlapped with backward;
\texttt{optimizer.py:668} --- ZeRO-1 sharding; \texttt{triton\_kernels.py:362,699}
--- fused RMSNorm, \texttt{:208,268} --- SwiGLU, \texttt{:1005,1166} --- RoPE,
\texttt{:112} --- fused cross-entropy; and the direct Transformer-Engine
attention path). These were rediscovered in the first nine steps of
Table~\ref{tab:framework-opt}.

\textbf{Vertical integration.} Components Megatron keeps as separate stages,
merged into one execution path. The optimizer tail is the clearest case:
Megatron runs gradient clipping, the Adam update, and the master-to-\texttt{bf16}
write-back as three passes over every parameter; the engine fuses all three
into one Triton kernel (\texttt{optimizer.py:357-385},
\texttt{triton\_kernels.py:415-446}).

\begin{lstlisting}
# optimizer.py:357 -- one fused_clip_adam_sync does clip + Adam + cast.
def fused_clip_adam_sync(state, fp32_grads, params, lr, clip_coeff):
    state.step_count += 1
    for name in state.param_names:
        g = fp32_grads[name].float()
        fused_adam_sync(g, master, exp_avg, exp_avg_sq, bf16_param,
                        lr=param_lr, step=state.step_count,
                        clip_coeff=clip_coeff, wd=wd)
\end{lstlisting}

\begin{lstlisting}
# triton_kernels.py:415 -- one pass per element: clip, Adam, cast.
g   = tl.load(grad_ptr+offs) * clip_coeff
m   = beta1*m + (1-beta1)*g
v   = beta2*v + (1-beta2)*g*g
m_hat = m / bias_correction1
v_hat = v / bias_correction2
update = m_hat / (tl.sqrt(v_hat) + eps) + wd * p
p_new = p - lr * update
tl.store(master_ptr+offs, p_new)
tl.store(bf16_ptr+offs, p_new.to(tl.bfloat16))
\end{lstlisting}

The same source also merges the per-layer graphs Megatron builds into one
captured CUDA graph over a whole microbatch's forward, loss, and backward
(\texttt{step\_graph.py:235-260}).

\textbf{Deep customization.} Points with no Megatron counterpart, where the
engine rewrites the execution path around the fixed model, shapes, and
parallel layout. The optimizer scalars and clipping coefficient are the
cornerstone: instead of computing the clip coefficient on the host from a
gradient-norm \texttt{.item()} round trip and passing the resulting Python
floats as kernel arguments (which would then be frozen at capture), they stay
on the device, in 1-element FP32 buffers the kernel reads via \texttt{tl.load}
(\texttt{optimizer.py:420-457}, \texttt{optimizer.py:460-524},
\texttt{triton\_kernels.py:525-566}).

\begin{lstlisting}
# optimizer.py:420 -- clip coefficient computed entirely on device.
def compute_clip_coeff_device(grad_norm_sq, clip_coeff_buf, max_norm, eps):
    grad_norm = grad_norm_sq.view(()).clamp(min=0.0).sqrt()
    coeff = (max_norm / (grad_norm + eps)).clamp(max=1.0)
    clip_coeff_buf.view(()).copy_(coeff)
    return clip_coeff_buf
\end{lstlisting}

\begin{lstlisting}
# triton_kernels.py:544 -- kernel reads per-step scalars from buffers.
lr   = tl.load(lr_ptr).to(tl.float32)
clip = tl.load(clip_coeff_ptr).to(tl.float32)
bc1  = tl.load(bc1_ptr).to(tl.float32)
bc2  = tl.load(bc2_ptr).to(tl.float32)
\end{lstlisting}

Because the values live in buffers rather than frozen launch arguments, the
host refreshes them once per step (\texttt{optimizer.py:492-524}) and the
replay stays correct; keeping the scalars on the device is what lets the step
avoid synchronizing with the host. The remaining deep-customization points
--- the loss all-reduce moved off the critical path, and the read-only,
layout-specific paths specialized to the fixed shapes --- complete the set.

\section{Ascend Framework-Level Optimization Trajectory}\label{app:ascend-framework-opt}

Table~\ref{tab:ascend-framework-opt} lists, in acceptance order, the gated
optimization rounds behind Table~\ref{tab:mfu-checkpoints} for the
MiniCPM5 130M engine, together with each round's own reported
$\mathrm{avg\ MFU(standard)}$ over its gate window. This is the per-round
detail behind that trajectory, analogous to Table~\ref{tab:framework-opt} on
H100. Unlike the H100 trajectory, which is governed by a single MFU target
throughout, the Ascend log applies two gates in sequence --- perf-bitwise
($\geq13\%$) then long-horizon ($\geq25\%$) --- so the phase column marks
which gate each round is scored against.

\begin{table*}[t]
\centering
\small
\caption{Gated optimization rounds applied to the MiniCPM5 130M engine
(18 layers), in acceptance order. MFU is that round's own
$\mathrm{avg\ MFU(standard)}$ over its gate window (Ascend 910, DP-only).}
\label{tab:ascend-framework-opt}
\begin{tabular}{@{}rllr@{}}
\toprule
\textbf{\#} & \textbf{Gate} & \textbf{Optimization} & \textbf{MFU (\%)} \\
\midrule
1  & perf-bitwise ($\geq13\%$) & Baseline (bitwise-preserving)              & 9.67  \\
2  & perf-bitwise ($\geq13\%$) & Mask/RoPE cache + foreach grad ops         & 9.73  \\
3  & perf-bitwise ($\geq13\%$) & Async grad-hash overlap                    & 15.27 \\
\midrule
4  & long-horizon ($\geq25\%$) & det-off + chunked CE (memory)              & 19.21 \\
5  & long-horizon ($\geq25\%$) & Dataloader prefetch                        & 20.10 \\
6  & long-horizon ($\geq25\%$) & CE upcast fused into log-softmax           & 20.05 \\
7  & long-horizon ($\geq25\%$) & Reuse forward log-softmax in backward      & 21.87 \\
8  & long-horizon ($\geq25\%$) & Phase-breakdown + host-sync lift           & 22.04 \\
9  & long-horizon ($\geq25\%$) & foreach-batched AdamW                      & 22.08 \\
10 & long-horizon ($\geq25\%$) & Fused NPU cross-entropy (FP32)             & 24.87 \\
11 & long-horizon ($\geq25\%$) & Defer grad-norm \texttt{.item()} off hot path & 24.88 \\
\bottomrule
\end{tabular}
\end{table*}

\begin{figure}[t]
\centering
\begin{subfigure}[b]{0.48\textwidth}
\includegraphics[width=\linewidth]{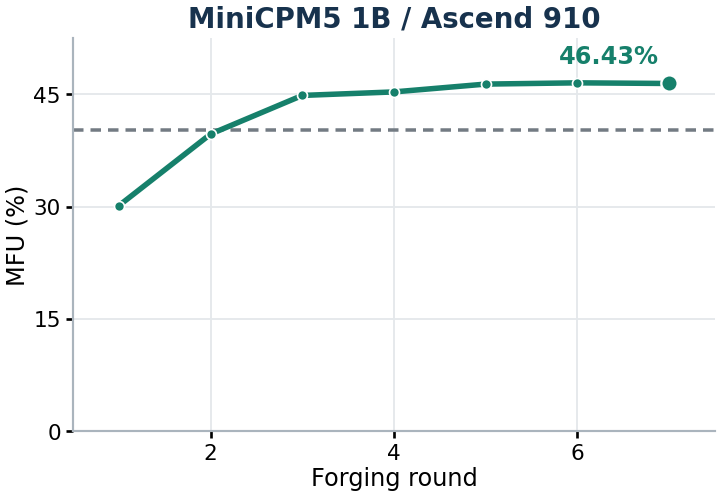}
\caption{MiniCPM5 1B}
\label{fig:mfu-ascend-1b}
\end{subfigure}\hfill
\begin{subfigure}[b]{0.48\textwidth}
\includegraphics[width=\linewidth]{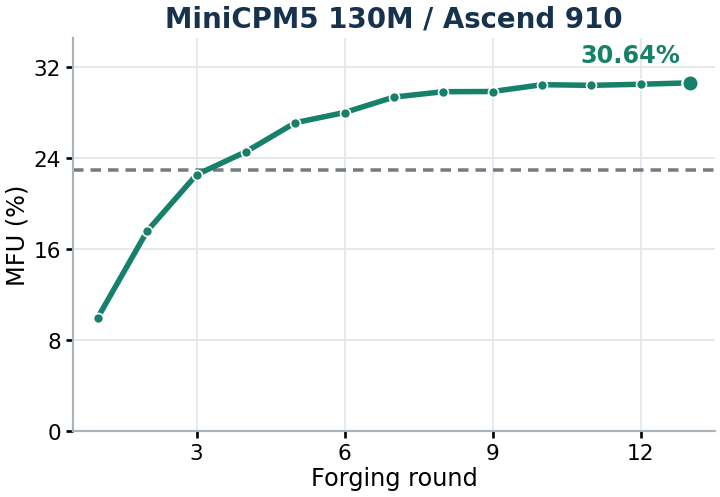}
\caption{MiniCPM5 130M}
\label{fig:mfu-ascend-130m}
\end{subfigure}
\caption{Per-setting MFU trajectories of the two Ascend 910 settings from
Table~\ref{tab:mfu-checkpoints}, each normalized to its own MindSpeed baseline
(dashed). The x-axis is the setting's $k$-th recorded forging round.}
\label{fig:mfu-panels-ascend}
\end{figure}

\section{Ascend Model Architectures}\label{app:ascend-models}

Table~\ref{tab:ascend-arch} lists the geometry for the two Ascend scenarios
referenced throughout this section, analogous to Table~\ref{tab:arch} on
H100. The 130M geometry (18 layers, \texttt{minicpm5\_130m}) is the one
the recorded forging rounds and the bit-for-bit anchors of
Table~\ref{tab:anchors} were measured on, chosen for cheaper
iteration; the 1B geometry (24 layers) is the engine pretrained end-to-end
in \S\ref{sec:e2e}.

\begin{table}[h]
\centering
\small
\caption{Bit-for-bit anchor counts captured from the golden reference. All
listed tensors must match exactly; MiniCPM5 130M uses blake2b equality.}
\label{tab:anchors}
\begin{tabular}{@{}lccc@{}}
\toprule
\textbf{Scenario} & \textbf{Forward} & \textbf{Backward} & \textbf{Total} \\
\midrule
MiniCPM4 0.5B & 183 & 339 & 522 \\
MiniCPM4 8B   & 164 & 195 & 359 \\
MiniCPM5 130M (Ascend) & 128 & 237 & 365 \\
\bottomrule
\end{tabular}
\end{table}

\begin{table*}[t]
\centering
\small
\caption{MiniCPM5 architecture configuration for the two Ascend scenarios.}
\label{tab:ascend-arch}
\begin{tabular}{@{}lcc@{}}
\toprule
\textbf{Hyperparameter} & \textbf{130M} & \textbf{1B (production)} \\
\midrule
Transformer layers                & 18   & 24   \\
Hidden size                       & 640  & 1536 \\
FFN intermediate size             & 1920 & 4608 \\
Attention heads                   & 10   & 16   \\
Key/value heads (GQA)             & 2    & 2    \\
Head dimension                    & 64   & 128  \\
Vocabulary size (as used in FLOPs) & 73448 & 130560 \\
Activation                        & SwiGLU & SwiGLU \\
Normalization                     & RMSNorm ($\epsilon{=}10^{-6}$) & RMSNorm \\
Position encoding                 & RoPE (base $10000$) & RoPE \\
Sequence length                   & 4096 & 4096 \\
Tied input/output embeddings      & Yes  & --- \\
$\mu$P (emb/depth/base-hidden)    & $12.0, 1.4, 256$ & $12.0, 1.4, 256$ \\
MTP / Eagle head                  & Off  & On (1 layer) \\
Training precision                & bfloat16 & bfloat16 \\
\bottomrule
\end{tabular}
\end{table*}

Two points are worth flagging rather than smoothing over. First, the 1B
scenario's FLOP accounting (Appendix~\ref{app:ascend-flops}) uses the full
MiniCPM5 tokenizer vocabulary ($130560$) because that is the field
\texttt{config.py}'s \texttt{\_global\_flops\_per\_token} reads, whereas the
130M scenario pads to $73448$ to match the pre-tokenized
data shards it trains against --- a real difference
in what the two model families' embedding/LM-head matrices are sized to, not
a reporting inconsistency. Second, the 1B config carries a one-layer Eagle
(MTP) head that the 130M scenario does not; its forward-plus-backward FLOPs
are the last row of Appendix~\ref{app:ascend-flops}'s table, bringing the 1B
total to $7.940$~GFLOP/tok, while the 130M total stays at the backbone-only
$1.070$.

\section{Ascend Model-FLOPs Utilization}\label{app:ascend-flops}

The Ascend engines report MFU against the same exact per-operator FLOP count
as \S\ref{app:flops}, not the closed-form approximation: both engines compute
the sum of attention projections, FFN (SwiGLU), attention core, and output head
per layer, forward, then multiply the total by $3$ for forward-plus-backward ---
the identical accounting \S\ref{app:flops} uses, down to the same $1{:}2$
forward-backward ratio applied to the output head. The 1B engine adds the forward
FLOPs of its one-layer Eagle (MTP) head to this sum when
\texttt{MTP\_ENABLED} is set, and we include that head as a separate row in
Table~\ref{tab:ascend-flops}; the 130M engine has no such head, so its total
is backbone-only. We reproduce the count here (Table~\ref{tab:ascend-flops})
rather than take the engines' self-reported denominators on faith: applying
the formula to the 0.5B/24-layer H100 geometry first reproduces
Table~\ref{tab:flops}'s $3.207$~GFLOP/tok exactly, which is what justifies
applying the same formula to the Ascend geometries of
Table~\ref{tab:ascend-arch}.

\begin{table*}[t]
\centering
\small
\caption{Per-token FLOP budget (forward${+}$backward) by operator class for
the two Ascend scenarios at sequence length $4096$, computed with the
\S\ref{app:flops} formula. The 1B total includes its one-layer Eagle (MTP)
head, shown as its own row; the 130M scenario trains without MTP, so that row
is not applicable there.}
\label{tab:ascend-flops}
\begin{tabular}{@{}lcccc@{}}
\toprule
& \multicolumn{2}{c}{\textbf{130M}} & \multicolumn{2}{c}{\textbf{1B (production)}} \\
\cmidrule(lr){2-3}\cmidrule(lr){4-5}
\textbf{Operator class} & \textbf{GFLOP/tok} & \textbf{Share} & \textbf{GFLOP/tok} & \textbf{Share} \\
\midrule
FFN GEMM (gate/up/down)          & $0.398$ & $37.2\%$ & $3.058$ & $38.5\%$ \\
Attention core                   & $0.283$ & $26.5\%$ & $1.208$ & $15.2\%$ \\
Output / CE head                 & $0.282$ & $26.4\%$ & $1.203$ & $15.2\%$ \\
Attn.\ projection GEMM (Q/K/V/O) & $0.106$ & $9.9\%$  & $1.019$ & $12.8\%$ \\
Eagle / MTP head                 & ---     & ---      & $1.452$ & $18.3\%$ \\
\midrule
\textbf{Total}                   & $1.070$ & $100\%$  & $7.940$ & $100\%$ \\
\bottomrule
\end{tabular}
\end{table*}

The attention core is a noticeably larger share on the Ascend 130M scenario
($26.5\%$) than on the H100 0.5B scenario ($18.8\%$,
Table~\ref{tab:flops}): the 130M model runs the same $4096$ sequence length
as H100 but with fewer, narrower heads ($10\times64$ versus $16\times64$), so
the FFN does not dominate the budget as heavily. On the 1B scenario the
attention share is lower still ($15.2\%$), because the one-layer Eagle (MTP)
head -- which repeats the backbone layer's attention, projection, and FFN
plus a second output head -- absorbs $18.3\%$ of the budget and dilutes the
backbone shares. This is the reason the fused-CE win of \S\ref{sec:throughput}
(which sits inside the output-head and attention-adjacent work) moves MFU by
a larger fraction on the 130M geometry than an equivalent fusion would on the
H100 0.5B geometry. Confirming that this numerator matches
\S\ref{app:flops}'s methodology narrows --- but does not close --- the
denominator question: what remains unverified is whether MindSpeed's own
\texttt{--log-throughput} uses the same per-operator count on its side of
Table~\ref{tab:mfu-checkpoints}, or a coarser approximation.

\section{Ascend Optimization Catalog}\label{app:ascend-opt-catalog}

This appendix supplements \S\ref{sec:throughput} with additional optimizations
recorded in the exported source but not detailed in the main text, in the
same two-tier shape as Appendix~\ref{app:opt-catalog}. Citations are
file:line references into the exported
\texttt{training\_engine\_tensor} packages.

\textbf{Framework level (goals MindSpeed shares, structure redrawn on
Ascend).}
\begin{itemize}[leftmargin=*,topsep=2pt,itemsep=3pt]
  \item \textbf{No autograd engine.} Both engines run their entire step under
    \texttt{torch.no\_grad()} with a hand-written static backward that
    replays the reference's own primitive backward ops in a fixed,
    pre-determined order (\texttt{backward.py:1-41}). MindSpeed's
    Megatron core builds and walks an autograd graph every step; the forged
    engines skip that construction and traversal entirely, at the cost of
    hand-maintaining the backward op sequence themselves.
  \item \textbf{Host-sync removal on the optimizer path.} The 130M engine
    lifts the optimizer step counter into a host Python int
    (\texttt{train\_loop.py:896-904}) and defers the grad-norm
    \texttt{.item()} call past the step-end device sync
    (\texttt{optimizer.py:207-240}), removing two CPU--GPU synchronization
    points from the hot path --- the same class of fix as H100's
    ``device-side gradient norm and clipping'' catalog item
    (Appendix~\ref{app:opt-catalog}), independently rediscovered on Ascend.
  \item \textbf{Chunked / streaming cross-entropy.} Beyond the fused FP32 CE
    operator described in \S\ref{sec:throughput}, the 1B engine carries a
    streaming CE path that folds the LM-head GEMM, the CE, and both
    gradient passes into one chunked loop that never materializes the full
    logits tensor at all --- not even a row-chunk at a time
    (\texttt{backward.py:276-368}). It is implemented and correctness-tested
    but not enabled in production (\texttt{train\_loop.py:1279},
    \texttt{\_streaming\_ce=False}); the in-place chunked path actually
    running in production (\S\ref{sec:throughput}) is the more conservative of
    the two.
\end{itemize}

\textbf{Framework level (no MindSpeed counterpart).}
\begin{itemize}[leftmargin=*,topsep=2pt,itemsep=3pt]
  \item \textbf{Weights as all-gather-buffer views.} The 1B engine's
    \texttt{bf16} parameter buffers are not copied out of the all-gather
    receive buffer after each collective; the weights are views into
    it, replaced in place on the next sync
    (\texttt{optimizer.py:1521-1552}), sparing roughly $1.8$\,GiB of
    duplicate \texttt{bf16} weight storage. This is the same zero-copy
    principle as H100's ``zero-copy gradient views'' catalog item
    (Appendix~\ref{app:opt-catalog}), applied on the parameter side instead
    of the gradient side.
  \item \textbf{Batched Newton--Schulz for Muon.} The 1B optimizer's Muon
    branch stacks same-shaped 2D parameters and runs Newton--Schulz
    iteration as batched \texttt{bmm} calls instead of one launch per
    parameter, cutting roughly $2500$ kernel launches to $60$
    (\texttt{optimizer.py:93-118}) --- an optimizer-level fusion with no
    Megatron-LM-side analog in \S\ref{sec:operators}, since the H100 engine
    does not run Muon.
\end{itemize}

\textbf{Kernel level (1B custom AscendC GEMM family; structural evidence
only --- see the header note on why no benchmark table is given here).}
\begin{itemize}[leftmargin=*,topsep=2pt,itemsep=3pt]
  \item \textbf{Per-shape specialized tiling.} Fifteen AscendC GEMM kernels
    (five projections $\times$ three directions: forward, \texttt{dgrad},
    \texttt{dwgrad}) hard-code $M/K/N$ as compile-time constants and tile
    dimensions sized to fill the 910's L1/L0C capacity exactly
    (\texttt{custom\_gemm.py:59-91}; e.g.\ \texttt{fc2\_fwd\_entry.cpp:16-33}
    fixes \texttt{AIC\_CNT=20}, \texttt{baseM=128}, \texttt{shareL1Size=
    516096}), the same per-shape specialization principle as H100's
    kernel-level catalog (Appendix~\ref{app:opt-catalog}), retargeted to a
    different vendor ISA.
  \item \textbf{Transpose elimination via \texttt{isTrans}.} The weight and
    gradient-output tensors are consumed in their natural physical layout
    rather than materialized-transposed first
    (\texttt{fc2\_fwd\_entry.cpp:81,114}), sparing a full-tensor HBM
    round-trip the vendor GEMM path would otherwise pay on the weight
    gradient.
  \item \textbf{Direct dispatch, bypassing the operator registry.} The
    custom kernels are invoked via \texttt{ctypes} against a compiled
    \texttt{.so}, passing raw \texttt{data\_ptr()}s and the current stream
    directly (\texttt{custom\_gemm.py:247-263,359-363}), skipping the
    \texttt{aten} dispatcher, \texttt{TensorIterator}, and the
    \texttt{torch\_npu} \texttt{aclnn} wrapper layer entirely.
\end{itemize}

%% file: figures/fig_attn_bwd.tex
\begin{figure*}[t]
\centering
\includegraphics[width=\linewidth]{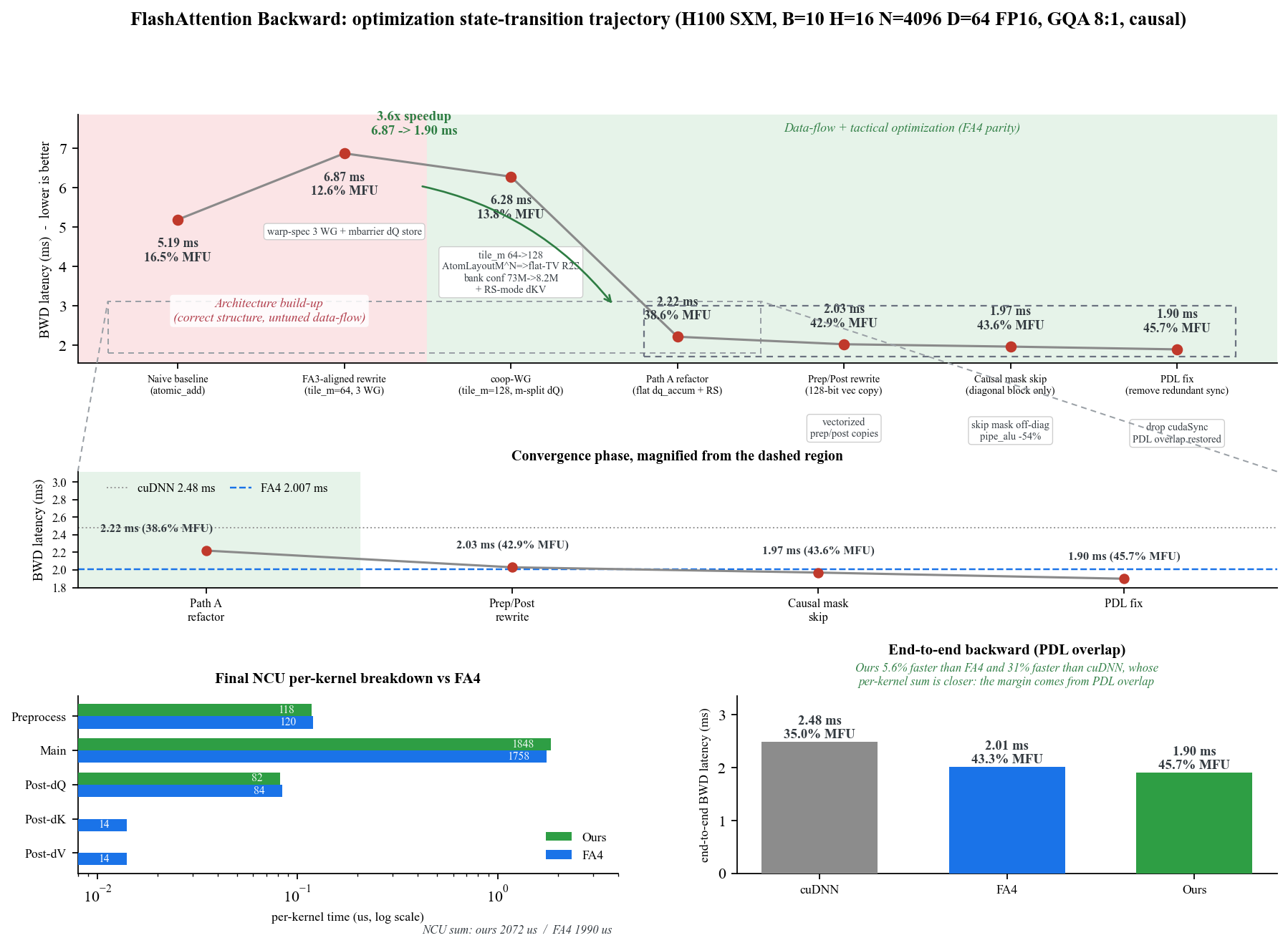}
\caption{Optimization trajectory of the FlashAttention backward kernel
($B{=}10$, $H{=}16$, $N{=}4096$, $D{=}64$, FP16, GQA $8{:}1$, causal; H100 SXM).
\textbf{Top:} backward latency (lower is better) across milestones. A structural
build-up phase (warp-specialized three-kernel pipeline, $128$-row $M$-tile,
double-buffered stages) is mutually dependent and regresses latency to
$6.3$--$6.9$\,ms before a $\mathrm{d}Q$ data-layout refactor (flat global buffer,
register-to-shared copy, register-sourced GEMM operands) drops the kernel to
$2.22$\,ms; three independent tactical steps---$128$-bit vectorized
preprocess/postprocess copies, diagonal-only causal masking, and removing a
stray synchronization that had disabled programmatic dependent launch---then
reach $1.90$\,ms ($45.7\%$ MFU). \textbf{Middle:} the four tactical steps
magnified from the dashed region of the top panel, on a $1.8$--$3.1$\,ms scale,
against the cuDNN and FlashAttention-4 end-to-end reference lines; the gains
are individually small ($2.22\to2.03\to1.97\to1.90$\,ms) and are what the top
panel cannot resolve. \textbf{Bottom left:} per-kernel NCU time
breakdown versus FlashAttention-4 (forged sum $2072$\,\textmu s vs.\ $1990$);
our per-kernel times are the slower ones, and the two post-$\mathrm{d}K$
kernels of FA-4 appear fused at $0.024$\,ms in ours. \textbf{Bottom right:}
end-to-end backward latency---against FlashAttention-4 ($2.007$\,ms,
$43.3\%$ MFU) the forged kernel is $5.6\%$ faster and against cuDNN
($2.483$\,ms, $35.0\%$ MFU) $31\%$ faster, despite the slower per-kernel sum:
the entire margin comes from programmatic dependent launch hiding $0.17$\,ms
across our four kernels, where FA-4's five show no measurable overlap.}
\label{fig:attn-bwd}
\end{figure*}